\documentclass[reprint,
 amsmath,amssymb,
 aps, prl]{revtex4-2}

\usepackage{graphicx}
\usepackage{dcolumn}
\usepackage{bm}
\usepackage{ulem}
\usepackage{xcolor}
\usepackage{slashed}
\usepackage{hyperref}

\newcommand{\cev}[1]{\reflectbox{\ensuremath{\vec{\reflectbox{\ensuremath{#1}}}}}}

\makeatletter
\g@addto@macro\normalsize{%
  \setlength{\abovedisplayskip}{12pt plus 1pt minus 9pt}%
  \setlength{\belowdisplayskip}{12pt plus 1pt minus 9pt}%
  \setlength{\abovedisplayshortskip}{4pt plus 3pt minus 4pt}%
  \setlength{\belowdisplayshortskip}{4pt plus 3pt minus 4pt}%
}
\makeatother

\AtBeginEnvironment{align}{%
  \setlength{\abovedisplayskip}{7pt plus 2pt minus 2pt}%
  \setlength{\belowdisplayskip}{7pt plus 2pt minus 2pt}%
  \setlength{\abovedisplayshortskip}{7pt plus 2pt minus 2pt}%
  \setlength{\belowdisplayshortskip}{7pt plus 2pt minus 2pt}%
}

\begin{document}

\title{Spontaneous Space-Time Parity Breaking Without Thermal Restoration}

\author{
Bilal Hawashin,$^{1,a}$
Michael M. Scherer,$^{1,b}$
Michael Smolkin,$^{2,c}$
and Lev Yung$^{2,d}$
}

\affiliation{$^{1}$Theoretische Physik III, Ruhr University Bochum, D-44801 Bochum, Germany\\
$^{2}$Racah Institute of Physics, The Hebrew University, Jerusalem 9190401, Israel}

\thanks{$^{a}$\href{mailto:hawashin@tp3.rub.de}{hawashin@tp3.rub.de}, 
        $^{b}$\href{mailto:scherer@tp3.rub.de}{scherer@tp3.rub.de}, 
        $^{c}$\href{mailto:michael.smolkin@mail.huji.ac.il}{michael.smolkin@mail.huji.ac.il}, 
        $^{d}$\href{mailto:lev.yung@mail.huji.ac.il}{lev.yung@mail.huji.ac.il}}

\date{\today}

\begin{abstract}
We construct an ultraviolet-complete, local, and unitary quantum field theory in 2+1 dimensions that exhibits spontaneous breaking of space-time parity, persisting to arbitrarily high temperatures.
The theory is defined by a renormalization group trajectory, triggered by a relevant deformation of a conformal field theory, consisting of a critical biconical vector model and a free massless Dirac fermion. 
This deformation couples the fermion to the scalar sector, generating a renormalization group flow that terminates at a nontrivial infrared fixed point described by a conformal Gross-Neveu-Yukawa model and a decoupled critical vector model. 
By construction, the quantum field theory is parity invariant at zero temperature. 
However, we show that at sufficiently high temperatures, parity symmetry is spontaneously broken and remains so even in the infinite-temperature limit. 
Our analysis relies on both, perturbative renormalization group techniques in $4\!-\!\epsilon$ dimensions and functional renormalization group techniques directly in 2+1 dimensions.
\end{abstract}

\maketitle

\section{Introduction}
According to the laws of thermodynamics, a system in thermal equilibrium minimizes its free energy. This principle governs the direction of spontaneous processes, driving the system toward equilibrium. At sufficiently high temperatures, entropy typically outweighs energy contributions in the free energy, favoring configurations with maximal disorder. Consequently, one generally expects any symmetry-breaking or ordered phases to be suppressed as temperature increases. However, this expectation, while common, is not a strict consequence of thermodynamic laws. While thermodynamics constrains the macroscopic behavior of equilibrium states, it does not uniquely determine the microscopic structure of the free energy landscape. As a result, certain systems may exhibit symmetry-breaking phases that persist as the temperature rises.

Indeed, Rochelle salt provides a striking counterexample to the common perception that increasing the temperature restores symmetry~\cite{landau1980statistical}. Below $-18^\circ\mathrm{C}$, it exists in an orthorhombic phase, which possesses a higher symmetry than the monoclinic phase that emerges above $-18^\circ\mathrm{C}$. Another noteworthy case is the Pomeranchuk effect in $^3\mathrm{He}$, where the solid phase exhibits higher entropy than the liquid state~\cite{Pomeranchuk:1950}. This counterintuitive behavior leads to solidification upon heating --- an effect that sharply contrasts with the typical thermodynamic response of most materials.

In relativistic quantum field theories~(QFTs), such examples are rare. The first known instance was constructed in a work by Steven~Weinberg in~\cite{weinberg1974gauge}, where it was argued that a four-dimensional model consisting of two weakly coupled massive scalar fields may exhibit spontaneous breaking of some of its internal symmetries at sufficiently high temperatures. It was shown that the broken phase persists to all orders in the weak coupling constant.

While these examples certainly challenge the idea of symmetry restoration by thermal fluctuations, they do not completely rule it out. Even before the temperature is high enough to destroy the crystalline structure of Rochelle salt, the system reaches a second Curie temperature at $+24^\circ\mathrm{C}$ and transitions into a phase with higher symmetry. The QFT example in~\cite{weinberg1974gauge} is ultraviolet~(UV) incomplete; therefore, the symmetry pattern becomes uncertain at finite temperatures corresponding to the strong coupling regime. In other words, these examples illustrate symmetry breaking within a finite window of intermediate temperatures, leaving the question of persistent symmetry breaking~(PSB) unresolved.
We note, however, that it has recently been shown that lattice models, i.e., models without explicit UV completion, exhibiting PSB can be constructed~\cite{Han:2025eiw}.

Recently, a class of relativistic, UV-complete scalar field theories in fewer than four space-time dimensions exhibiting PSB was constructed~\cite{Chai:2020zgq,Chai:2020onq,Chai:2020hnu,Chai:2021djc,Chai:2021tpt,Liendo:2022bmv,Hawashin:2024dpp,Komargodski:2024zmt,Han:2025eiw}. The UV-completeness of these models is ensured by their conformal invariance. However, many of them suffer from certain limitations. For example, the models in~\cite{Chai:2021djc,Chai:2021tpt} are non-local, while the analysis in~\cite{Chai:2020zgq,Chai:2020onq} relies on the $\epsilon$~expansion, which raises concerns about unitarity~\cite{Hogervorst:2015akt} unless $\epsilon$ is extended to an order-one value to extrapolate the model to 2+1 dimensions. Although such extrapolation is commonly used in the study of critical exponents, it cannot serve as a rigorous demonstration of PSB in 2+1 dimensions. A direct calculation in 2+1 dimensions is therefore essential. Such a calculation was carried out in~\cite{Hawashin:2024dpp}, where the authors employed a truncated functional renormalization group~(FRG) approach, corroborating the indications for persistent breaking of a discrete symmetry in a 2+1 dimensional, local, UV-complete, and unitary QFT. Furthermore, \cite{Komargodski:2024zmt} proposed a class of well-defined and analytically tractable conformal field theories~(CFTs) in 2+1 dimensions where PSB of internal discrete symmetries can be demonstrated explicitly.

In four dimensions, the situation is technically more involved. Despite extensive study, no local, UV-complete, unitary, interacting QFT in 3+1 dimensions built solely from scalars has been found. Perturbatively tractable examples of interacting CFTs in four dimensions are limited to the Banks–Zaks–Caswell type \cite{Belavin:1974gu,Caswell:1974gg,Banks:1981nn}. These theories necessarily involve gauge fields and fermions — not scalars alone. The possibility of PSB in CFTs of this type has been illustrated in \cite{Chaudhuri:2020xxb,Bajc:2020gpa,Chaudhuri:2021dsq}. However, the analysis in these works is restricted to the planar limit, and no extension beyond it has been achieved to date.

Here, we report the first known instance of persistent breaking of a space-time symmetry --- parity in 2+1 dimensions. Unlike previous constructions, our model includes a Dirac field and explicitly breaks conformal symmetry. It defines a local, UV-complete, relativistic QFT realized as a renormalization group~(RG) trajectory connecting two fixed points in 2+1 dimensions. In the deep infrared~(IR), the flow terminates at a fixed point consisting of a decoupled pair of well-known CFTs: the Gross–Neveu–Yukawa (GNY) model and the critical vector model~\cite{Moshe:2003xn}. On the UV end, it originates from a CFT describing a free massless Dirac field decoupled from the critical biconical vector model, studied via the $\epsilon$-expansion~\cite{Fisher_mcl:1974,Vicari:2003,Chai:2020zgq,Chai:2020onq} and FRG~\cite{Eichhorn_mcl2013,Hawashin:2024dpp}. The RG flow connecting the UV and IR fixed points is initiated by a relevant Yukawa-type deformation that couples the Dirac fermion to the biconical vector model. Using FRG and the $\epsilon$-expansion, we show that the $2+1$-dimensional QFT defined by this RG flow exhibits persistent parity breaking at arbitrarily high temperatures.

The significance of parity violation spans a wide range of physical systems, including the Standard Model, the quantum Hall effect, topological order, anyonic excitations, topological insulators, and spin liquids. Moreover, PSB of parity may have played a role in particle physics and early universe cosmology; see, e.g., \cite{Meade:2018saz,Bai:2021hfb,Ramazanov:2020ajq,Ramazanov:2021eya} and references therein. It would be interesting to explore whether our model — or suitable extensions — can contribute to potential applications in these directions.

In 2+1 dimensions, the search for PSB is restricted to discrete symmetries by the Coleman–Mermin–Wagner–Hohenberg theorem~\cite{PhysRevLett.17.1133,PhysRev.158.383,Coleman:1973ci}. In contrast, no analogous no-go theorem exists in 3+1 dimensions, allowing, in principle, for the persistent breaking of continuous symmetries. Identifying robust examples of 3+1-dimensional QFTs with a finite number of fields that exhibit persistent order remains an important open challenge.

Ultimately, a deeper understanding of the mechanisms underlying PSB is essential for realizing practical applications. An initial step in this direction was taken in~\cite{Han:2025eiw}, which investigated the phenomenon in a variety of models, including lattice-based constructions. Further insights may be gained through the AdS/CFT correspondence, with interesting results reported in~\cite{Buchel:2020thm,Buchel:2020xdk,Buchel:2020jfs,Buchel:2021ead,Buchel:2022zxl,Buchel:2023zpe,Buchel:2025cve}.

\section{Model}
Consider a $d$-dimensional Euclidean field theory involving a real pseudo-scalar field~$\chi$, a Dirac fermion~$\psi_i$ transforming in the fundamental representation of~$U(N_f)$, and a real scalar field~$\phi =(\phi_1,\dots,\phi_{N_1})$ transforming as a vector under~$O(N_1)$. The action is given by~\footnote{For brevity, we omit the explicit flavor indices of $\phi$ and $\psi$; summation over flavor contractions is implied throughout.}
\begin{multline}
    S = \int d^d x \left\{\frac{1}{2}(\partial_\mu\phi)^2 + \frac{1}{2} (\partial_{\mu} \chi)^2 + \bar{\psi} \slashed{\partial} \psi + \right.
    \\
    \left. +\frac{\lambda_{\phi}\mu^{\epsilon}}{8} (\phi^2)^2\! +\! \frac{\lambda_{\chi}\mu^{\epsilon}}{8} \chi^4\! +\! \frac{\lambda_{\phi\chi}\mu^{\epsilon}}{4} \phi^2 \chi^2 \!+\! h\mu^{\epsilon/2}\chi \bar{\psi} \psi \right\},
    \label{action}
\end{multline}
where $d=4-\epsilon<4$ and $\mu$ is a mass scale, so that all dimensionless couplings are relevant deformations of the Gaussian fixed point. In $d=2+1$ dimensions, this model possesses an $O(N_1) \times U(N_f) \times \mathcal{P}$ symmetry, where $\mathcal{P}$ stands for parity,
\begin{equation}
\mathcal{P}: \,\,\, 
x_1\to -x_1,\,\,\, \chi\to -\chi, \,\,\, \psi\to \gamma^1\psi, \,\,\,
\bar\psi\to-\bar\psi\gamma^1.
\end{equation} 
Fractional dimensions with $0<\epsilon<1$ are treated via analytic continuation. For later convenience, we define $N_2 = d_\gamma N_f$, where $d_\gamma$ is the dimension of the spinor representation.

By construction, this field theory admits a rich structure of multi-critical fixed points. In $d = 2+1$ dimensions, these fixed points are strongly coupled. Some of the associated CFTs are well-known and extensively studied in the literature. For example, setting $\lambda_{\phi\chi} = 0$ decouples the field $\phi$ from the other two fields, $\chi$ and $\psi$. The resulting RG flow terminates at a nontrivial IR fixed point described by a decoupled product of two familiar CFTs: the GNY model and the critical vector model \cite{Moshe:2003xn}. This particular fixed point plays an important role in our construction; we will refer to it as $\text{CFT}_{\text{IR}}$.

The existence of critical GNY and scalar vector models in $d=2+1$ dimensions has been confirmed by various complementary methods, including the large-$N$ expansion, FRG, $\epsilon$-expansion, conformal bootstrap, as well as lattice and Monte Carlo simulations, see e.g., \cite{Zinn-Justin:1991ksq,Hands:1992be,Karkkainen:1993ef,doi:10.1142/S0217751X94000285,doi:10.1142/S0217751X94000340,Hasenbusch:2010hkh,Kos:2013tga,Janssen:2014gea,PhysRevB.94.245102,PhysRevB.98.125109,Erramilli:2022kgp,PhysRevB.108.L121112}.

Another interesting CFT in \eqref{action} is located on the hypersurface in coupling space defined by $h = 0$. On this hypersurface, the Dirac field decouples from the scalar sector, while the remaining couplings can be tuned to the critical values corresponding to the conformal biconical vector model with $\lambda^*_{\phi\chi} < 0$ for sufficiently large~$N_1$, which has a stable potential. This CFT has been studied using large-$N$ and $\epsilon$-expansion techniques~\cite{Fisher_mcl:1974,Vicari:2003,Chai:2020zgq,Chai:2020onq,Chai:2020hnu}, as well as via the FRG approach directly in $d=2+1$~\cite{Eichhorn_mcl2013, Hawashin:2024dpp}. We will henceforth refer to the direct product of this CFT and a free massless Dirac fermion as $\text{CFT}_{\text{UV}}$.

The QFT of interest is defined by deforming $\text{CFT}_{\text{UV}}$ with a Yukawa interaction controlled by the coupling~$h$ in Eq.~\eqref{action}. The action of the resulting theory is given by
\begin{equation}
    S'=S_\text{UV} +\int d^dx \bar{\psi} \slashed{\partial} \psi + h \,\mu^{\epsilon/2}\int d^dx \, \chi \bar{\psi} \psi ~,
    \label{Smodel}
\end{equation} 
where $S_\text{UV}$ denotes the action of the conformal biconical vector model, characterized by $\lambda^*_{\phi\chi} < 0$. As we will argue below, this deformation is relevant and induces an RG flow that drives the theory toward $\text{CFT}_{\text{IR}}$ in the infrared limit. We note that in the $\epsilon$-expansion approach, the RG relevant mass terms are already chosen at their critical values, and no additional fine-tuning is required. This will be different in the FRG approach. 

The model is strongly coupled in $2+1$ dimensions. Therefore, in the next section, we employ numerical FRG methods to demonstrate that $S'$ flows to $\text{CFT}_{\text{IR}}$ in the deep infrared. In the remainder of this section, we assume $\epsilon \ll 1$ and establish this connection within the framework of the $\epsilon$-expansion.

\begin{figure}[htbp]
  \centering
  \includegraphics[
    width=0.48\textwidth,         
    keepaspectratio,              
    page=1                        
  ]{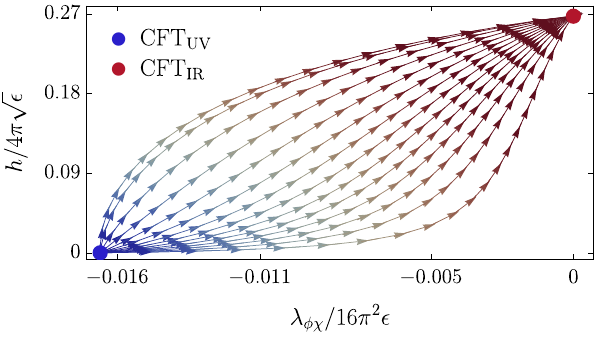}
  \caption{RG flow from the $\epsilon$-expansion for $N_1=100$, $N_2=8$, shown in the $h$-$\lambda_{\phi\chi}$ coupling plane. Within this approximation, most deformations of CFT$_\text{UV}$ with $h\neq 0$ flow to CFT$_\text{IR}$, which serves as an RG sink with no relevant directions. Trajectories flow into CFT$_\text{IR}$ but not out, i.e., it is an IR stable fixed point.}
  \label{fig:epsRG}
\end{figure}

The one-loop beta functions are derived in \cite{supplementary},
\begin{align} 
    \beta_{\lambda_\phi} &= - \epsilon \lambda_\phi + {N_1+8\over 16 \pi^2} \lambda_\phi^2
    + {\lambda_{\phi\chi}^2\over 16 \pi^2},\nonumber 
     \\
    \beta_{\lambda_{\phi\chi}} &= - \epsilon \lambda_{\phi\chi} + {N_2\over 16\pi^2} \lambda_{\phi\chi} h^2
    +{3 \over 16\pi^2}\lambda_\chi \lambda_{\phi\chi} +{\lambda_{\phi\chi}^2\over 4\pi^2}
    \nonumber\\
    &+{N_1+2\over 16\pi^2} \lambda_{\phi} \lambda_{\phi\chi} , \label{eq: betaeps}\\
\beta_{\lambda_\chi} &= -\epsilon\lambda_\chi +\frac{1}{8\pi^2}\Big(\frac{9}{2}\lambda_\chi^2 + {N_1\over 2}\lambda_{\phi\chi}^2 + 
         N_2\lambda_\chi h^2-2N_2h^4\Big) , \nonumber\\
  \beta_{h^2} &= -\epsilon h^2+\frac{N_2+6}{16\pi^2}h^4 \nonumber. 
\end{align}
For $N_1 \geq 11$, one can verify the existence of a weakly coupled fixed point characterized by $h_* = 0$ and $\lambda^*_{\phi\chi} < 0$, which we denote as CFT$_{\text{UV}}$. At this point, the anomalous dimension of $\chi$ is of order $\mathcal{O}(\epsilon^2)$, indicating that the Yukawa deformation $\sim h$ of CFT$_\text{UV}$ is relevant. It induces an RG flow that drives the theory toward the $\text{CFT}_{\text{IR}}$ in the infrared limit. Fig. \ref{fig:epsRG} illustrates typical RG flows associated with Eqs.~\eqref{Smodel},~\eqref{eq: betaeps}. 
In the next section, we confirm this observation directly in $d=2+1$.

\section{FRG analysis in 2+1D}
Consider a microscopic action $S_\Lambda[\Phi]$, which is a functional of a collective field variable $\Phi(x)$ defined at some UV scale $\Lambda$. Within the FRG \cite{Wetterich:1992yh, Berges:2000ew}, we define the \textit{effective average action} $\Gamma_k[\Phi]$ by the path integral
\begin{equation}
        e^{-\Gamma_k[\Phi]} = \int D\hat{\Phi} e^{-S_\Lambda[\Phi +\hat{\Phi}] + \int d^d x\, \frac{\delta \Gamma_k[\Phi]}{\delta \Phi} \hat{\Phi} - \Delta S_k[\hat{\Phi}]}.
\end{equation}
In the above, $\Phi(x)$ is a classical field configuration and $\hat{\Phi} (x)$ is the quantum field integrated over in the path integral. We further introduced the regulator insertion $\Delta S_k$ acting as an IR regulator, i.e., in the path integral, it suppresses modes below the energy scale $k$. Thus, the average action $\Gamma_k[\Phi]$ contains all the quantum fluctuations of the theory above the energy scale $k \in [0, \Lambda]$, and can be viewed as an effective action for field averages over $k^{-d}$ volume.

In the case of our field content $\Phi = (\phi, \chi, \psi, \bar{\psi})$, we choose the regulator insertion bilinear in the fields,
\begin{multline}
\Delta S_k[\Phi] = \int \frac{d^d q}{(2\pi)^d}\left\{\frac{1}{2} \phi(-q) R^{\phi}_{k}(q) \phi(q) +\right.\\[4pt]
\left. + \frac{1}{2}\chi(-q) R^{\chi}_{k}(q) \chi(q) + \bar{\psi}(q) R^\psi_{k}(q) \psi(q) \right\}, 
\end{multline}
where $R_k(q)$ is large for $q<k$ and vanishes for $q \gtrsim k$. For example, choosing a sharp IR regulator makes the relation of $\Gamma_{k\to \Lambda}$ to the microscopic action obvious and shows that $\Gamma_{k}$ interpolates between the microscopic action in the UV and the full quantum effective action in the IR, i.e., $\Gamma_{k=\Lambda}[\Phi] = S_\Lambda[\Phi]$,\, $ \Gamma_{k=0}[\Phi] = \Gamma[\Phi]$~\cite{Pawlowski:2005xe}. 

The dependence of $\Gamma_k$ on $k$ is governed by the Wetterich equation \cite{Wetterich:1992yh}
\begin{equation} \label{eq: Wetterich}
    \partial_t \Gamma_k [\Phi] = \frac{1}{2} \text{STr} \left(\left[\Gamma^{(2)}_k[\Phi]+R_k\right]^{-1} \partial_t  R_k\right),
\end{equation}
where $t = \ln \frac{k}{\Lambda}$ is the FRG time, the supertrace is taken over space-time
degrees of freedom, field and spinor indices with a minus sign for the fermionic sector. The inverse average propagator $\Gamma^{(2)}_k$ is the second functional derivative of $\Gamma_k$
\begin{equation}
\left(\Gamma^{(2)}_k[\Phi]\right)_{p,q;\,ij} = \frac{\vec{\delta}}{\delta \Phi_{i}(-p)} \Gamma[\Phi] \frac{\cev{\delta}}{\delta \Phi_{j} (q)}, 
\end{equation}
with $\Phi(p) = (\phi(p), \chi(p), \psi(p),\bar{\psi}(-p))$.  The regulator matrix $R_k$ is the second functional derivative of $\Delta S_k[\Phi]$ with respect to the field arguments. The Wetterich equation~\eqref{eq: Wetterich} is an exact identity. However, in general cases, it can not be solved exactly, and an appropriate truncation scheme for the average action is needed to limit the number of coupled differential equations.

We choose our truncation scheme to be the leading order in the derivative expansion, known as the extended local potential approximation (LPA$'$),
\begin{multline} \label{eq: truncation}
    \Gamma_k[\Phi] = \int d^d x \left\{\frac{Z_{\phi,k}}{2}(\partial\phi)^2 + \frac{Z_{\chi,k}}{2} (\partial \chi)^2 + U_k[\rho_\phi, \rho_\chi]\right.\\ + Z_{\psi,k}\bar{\psi}\slashed{\partial} \psi  + h_k\chi \bar{\psi} \psi \Bigr\}.
\end{multline}
This truncation includes the scale-dependent wave-function renormalizations $Z_{ \phi/\chi/\psi,k}$, scale-dependent Yukawa coupling $h_k$, and scale-dependent effective potential $U_k$ as a function of global symmetry invariants $\rho_{\phi} = \frac{1}{2} \phi_i \phi_i$ and $\rho_{\chi} = \frac{1}{2} \chi^2$. For brevity, we omit the $k$ subscripts further on. Close to four dimensions, the above ansatz is able to reproduce the one-loop results from the $\epsilon$-expansion for certain regulator choices, see, e.g.,~\cite{Zambelli:2021}. This scheme was successfully used in many works, see e.g., \cite{Hofling:2002hj,Scherer:2012fjq, janssen2014antiferromagnetic, classen2017fluctuation, Hawashin:2024dpp, Tolosa-Simeon:2025fot}.

We define the dimensionless effective potential $u\equiv k^{-d}U$, and its derivatives $ u^{(l,m)} \equiv \partial_{\bar{\rho}_\phi}^l \partial_{\bar{\rho}_\chi}^m u$ in terms of the dimensionless and renormalized $\bar{\rho}_{\phi/\chi} \equiv Z_{\phi/\chi} k^{2-d}\rho_{\phi/\chi}$. Likewise, dimensionless and renormalized Yukawa coupling is given by $\bar{h} \equiv k^{d/2-2} Z_{\chi}^{-1/2} Z_{\psi}^{-1} h$. 

Employing the linear regulator \cite{Litim:2001up}, we deduce from (\ref{eq: Wetterich}) the FRG flow of all the couplings of the truncated average action (\ref{eq: truncation}). Finite temperature is introduced by compactifying the time direction on a circle of radius $\beta = \frac{1}{T}$. The flow of the dimensionless effective potential at finite temperature reads 
\begin{flalign} \label{eq: uflow}
    \partial_t u = &-d u + (d-2 + \eta_\phi) \bar{\rho}_\phi u^{(1,0)} + (d-2 + \eta_\chi) \bar{\rho}_\chi u^{(0,1)} \notag\\ &+ I_R(\omega_{\chi},\omega_\phi,\omega_{\phi \chi}) S_\phi (\tau) + (N_1-1) I_G(u^{(1,0)})  S_\phi (\tau) \notag\\ &+ I_R(\omega_{\phi},\omega_\chi,\omega_{\phi \chi}) S_\chi (\tau) - N_2 I_F (\omega_\psi) S_\psi (\tau),
\end{flalign}
where $\eta_i = - \partial_t \ln Z_{i}$ are anomalous dimensions, $S_i (\tau) = s_i (\tau) - \frac{\eta_i}{d+1+\delta_{i,B}} \hat{s}_i(\tau)$ with $\tau \equiv 2\pi T/k$, $i \in \{\phi,\chi,\psi\}$, $B = \{\phi, \chi\}$. Thermal functions $s_i$, $\hat{s}_i$ are normalized to unity at $T = 0$. Arguments of the $I_{R,G,F}$ threshold functions read: $\omega_{\phi} \equiv u_k^{(1,0)} + 2\bar{\rho}_\phi u_k^{(2,0)}$, $\omega_{\chi} \equiv u_k^{(0,1)} + 2\bar{\rho}_\chi u_k^{(0,2)}$, $\omega_{\phi \chi} \equiv 2 \sqrt{\bar{\rho}_\phi \bar{\rho}_\chi} u^{(1,1)}$, $\quad \omega_\psi = 2\bar{\rho}_\chi \bar{h}^2$. The first line of (\ref{eq: uflow})  arises from the classical and anomalous scalings, while subsequent lines represent contributions from fluctuations of the radial $\phi$-mode, $N_1-1$ transverse $\phi$-modes, the radial $\chi$-mode, and fermionic modes, respectively.

The flow of the Yukawa coupling at finite temperature reads
\begin{multline} \label{eq: hflow}
\partial_t \bar{h}^2 = (d-4 + \eta_\chi + 2\eta_\psi) \bar{h}^2  +  8 v_d \left(
\bar{h}^3 l_{(1,1)}^{\chi\chi} - 2 \omega_\psi \bar{h}^3 l_{(2,1)}^{\chi\chi} \right.\\\left.- \bar{h} \sqrt{\omega_\psi} \left(
\omega_{\chi\chi\chi} l_{(1,2)}^{\chi\chi} + \omega_{\chi\phi\phi} l_{(1,2)}^{\phi\phi} - 2 \omega_{\chi\chi\phi} l^{\phi\chi}
\right)
\right),
\end{multline}
where $v_d^{-1} = 2^{d+1} \pi^{\frac{d}{2}}\Gamma(\frac{d}{2})$, the $l$ threshold functions are functions of $d$, $\tau$, $\eta_i$ and $\omega$'s, $\omega_{ijk}$ are algebraic–differential functions of $u$. For all the explicit expressions omitted here, including the anomalous dimensions, consult the Supplementary Materials \cite{supplementary}.

We expand the effective dimensionless potential locally around its running minimum $(\kappa_\phi, \kappa_\chi) := \left.(\rho_\phi, \rho_\chi)\right|_{\min U_k}$. In the IR limit $k\rightarrow 0$, $\kappa_\phi$ and $\kappa_\chi$ are the vacuum expectation values $\frac{1}{2}\langle\phi\rangle^2$ and $\frac{1}{2}\langle\chi\rangle^2$ respectively. At a given FRG scale $k$ we distinguish four different regimes depending on the location of the $U_k$ minima: (1) SYM-SYM: $\kappa_\phi = \kappa_\chi = 0$, (2) SYM-SSB: $\kappa_\phi = 0,\, \kappa_\chi > 0$, (3) SSB-SYM: $\kappa_\phi > 0,\, \kappa_\chi = 0$, (4) SSB-SSB: $\kappa_\phi > 0$, $\kappa_\chi > 0$. At $k = 0$ these regimes respectively correspond to the $O(N_1)\times \mathbb{Z}_2$, $O(N_1)$, $O(N_1-1)\times \mathbb{Z}_2$ and $O(N_1-1)$ unbroken symmetries of the theory in the bosonic sector.

The effective potential is expanded up to total polynomial order $n_{\text{max}}$, known as the LPA$'(2n_\text{max})$ truncation,
\begin{equation} \label{eq: eff_potential}
    u(\bar{\rho}_\phi, \bar{\rho}_\chi) = \sum_{n+m\geq 1}^{n+m \leq n_{\text{max}}} \frac{\bar{\lambda}_{nm}}{n! m!} (\bar{\rho}_\phi-\bar{\kappa}_\phi)^n (\bar{\rho}_\chi-\bar{\kappa}_\chi)^m,
\end{equation}
with $\bar{\lambda}_{1,0} \equiv \bar{m}^2_\phi \geq 0$ if $\bar{\kappa}_\phi = 0$ (SYM), and $\bar{\lambda}_{1,0} = 0$ if $\bar{\kappa}_\phi > 0$ (SSB); similarly for $\bar{\lambda}_{0,1}$ and $\bar{\kappa}_\chi$. For each of the four regimes, using (\ref{eq: uflow}), one gets the $\beta^{\text{FRG}}$-functions of the $\bar{\kappa}_\phi$, $\bar{\kappa}_\chi$ and $\bar{\lambda}_{nm}$ couplings \cite{supplementary}.

\section{2+1D Flow}

Model (\ref{action}) admits two interacting CFTs, denoted by CFT$_\text{UV}$ and CFT$_\text{IR}$. CFT$_\text{UV}$ is the product of the well-established biconical fixed point of the $O(N_1)\times \mathbb{Z}_2$ model and a decoupled free massless fermion. CFT$_\text{IR}$ is the decoupled product of the critical $O(N_1)$ vector model and the critical Gross-Neveu-Yukawa model, both well-known in the literature. In the following, we restrict to $d = 3$, $T = 0$, $N_1 = 100$, $N_2 = 8$ with all the computations performed within the non-perturbative FRG framework in the LPA$'8$ truncation~\eqref{eq: truncation}, \eqref{eq: eff_potential}. Large $N_1$ provides
rapid convergence of the biconical and vector models in the LPA$'$ and LPA truncations \cite{Hawashin:2024dpp, Eichhorn_mcl2013, DAttanasio:1997yph}.
We further chose $N_2 = 8$ for better comparability with literature and its relevance for condensed matter systems, see, e.g.,~\cite{herbut2009theory}.
In particular, for the chosen $N_1$ and $N_2$, our values of critical exponents and anomalous dimensions for the biconical, vector, and GNY critical models are in reasonable agreement with recent literature~\cite{supplementary}. Table \ref{tab: CFTs} summarizes a number of critical coupling values and positive scaling exponents of the two CFTs. Higher order truncations up to LPA$'12$ do not yield any significant improvement in numerical precision. Scaling exponents $\theta_i$'s are minus the eigenspectrum of the stability matrix $B_{ij}= \left.\partial\beta_i/\partial \lambda_j\right|_{\lambda=\lambda^*}$ evaluated at the fixed point. We denote $g_i$ to be the corresponding eigendirection in the space of couplings, and $\Delta g_i$ as the size of the deformation from the fixed point along this direction.
\begin{table}[h]
\centering
\begin{tabular}{l|ccccccccccc}
\hline\hline
 \tiny{SSB-SSB}& \(\kappa_\phi\) & \(\kappa_\chi\) & \(\lambda_\phi\) & \(\lambda_\chi\) & \(\lambda_{\phi\chi}\) & $h$ & \(\theta_1\) & \(\theta_2\) & \(\theta_3\) & $\theta_h$ \\
\hline
CFT\(_\mathrm{UV}\) & 1.93             & 0.26             & 0.29              & 0.31              & –0.29                  & 0          & 1.99          & 1.01     &  0.95   &  0.50\\
\hline\hline
\tiny{SSB-SYM} & \(\kappa_\phi\) & $\lambda_\phi$ & \multicolumn{1}{c|}{$\theta$} & \(m_\chi^2\) &  \(\lambda_\chi\) & $h$ & \multicolumn{1}{c|}{$\theta$} & \(\lambda_{\phi\chi}\) &  &  &  \\
 \hline
CFT\(_\mathrm{IR}\) & 1.68             &  0.29 & \multicolumn{1}{c|}{1.01}          & 0.40              & 5.42       & 1.51      & \multicolumn{1}{c|}{0.98}  & 0        &      &     & \\
\hline\hline
\end{tabular}
\caption{Fixed-point data of CFT$_\text{UV}$ and CFT$_\text{IR}$. Displayed are the critical couplings and positive scaling exponents $\theta_i$'s, in particular, $\theta_h$ corresponds to the pure Yukawa $h$-direction. CFT$_\text{IR}$ data is split into the two decoupled sectors for clarity. CFT$_\text{UV}$ is in the SSB-SSB regime, where both $\kappa_{\phi/\chi} > 0$, while $m_{\phi/\chi}^2 = 0$ and CFT$_\text{IR}$ is in the SSB-SYM regime where $\kappa_{\phi},\,m_\chi^2 > 0$ and $\kappa_{\chi}=m_\phi^2 = 0$.}
\label{tab: CFTs}
\end{table}

\begin{figure}[htbp]
  \centering 
    \includegraphics[
    width=0.47\textwidth,        
    keepaspectratio,
    page=1
  ]{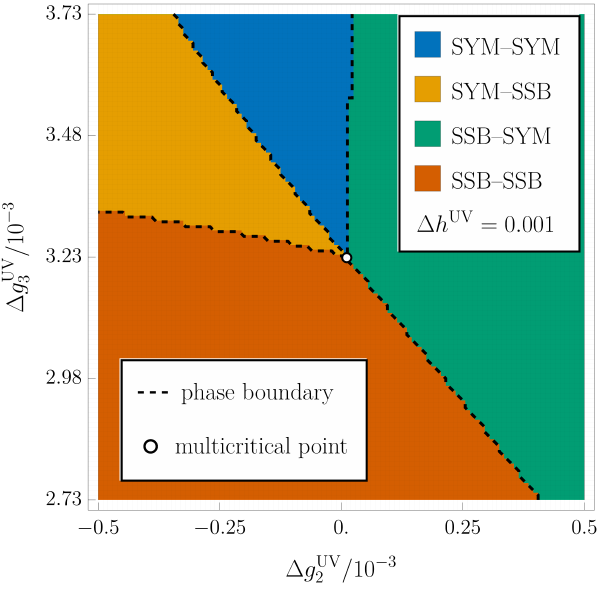}
  \caption{Infrared phases of deformed CFT$_\text{UV}$. The relevant deformation $\Delta h$ is fixed, while $\Delta g_2$ and $\Delta g_3$ are varied. The deformation flowing to the multicritical point can be reduced by a factor of 10 and still flow to the vicinity of the multicritical point, confirming the smallness of CFT$_\text{UV}$ deformation.
  } 
  \label{fig: frg_phase_diagram}
\end{figure}

\begin{figure}[htbp]
  \includegraphics[
    width=0.49\textwidth,
    keepaspectratio,
    page=1
  ]{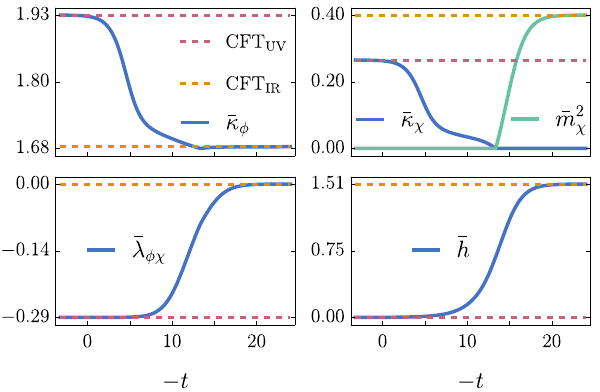}
  \caption{FRG flow connecting CFT$_\text{UV}$ and CFT$_\text{IR}$. Dashed lines correspond to the critical coupling values. At $t = 0$ the deformation from CFT$_\text{UV}$ is $\Delta g_2 = 0.0000112437360545$, $\Delta g_3 = 0.0032287264705453$, $\Delta h = 0.001$. At $t \approx -13.4$, the flow switches from the SSB-SSB to the SSB-SYM regime, consistent with the fixed point data from Table \ref{tab: CFTs}.} 
  \label{fig: frg_flow}
\end{figure}

We choose to deform CFT$_\text{UV}$ along its three relevant directions $g_2$, $g_3$, and $h$, triggering an RG flow that drives the theory away from CFT$_\text{UV}$ toward a different point in theory space. Figure \ref{fig: frg_phase_diagram} shows the IR phases of the deformed theories. The phase diagram clearly shows a multicritical point where all four phases meet. According to the theory of critical phenomena, this point is a CFT. Indeed, it corresponds to CFT$_\text{IR}$, which is a decoupled product of two critical models. By tuning their respective relevant deformations, one is undergoing the spontaneous symmetry-breaking transition in the $\phi$ field, and the other in the $\chi$ field. Figure \ref{fig: frg_flow} shows a concrete example of the RG flow between CFT$_\text{UV}$ and CFT$_\text{IR}$. 

Our QFT model is defined by specifying an energy scale $\Lambda$ and selecting an interior point along the RG trajectory connecting two CFTs, thereby fixing dimensionless couplings at this scale. Within the $\epsilon$-expansion, turning on solely the relevant coupling $h$ suffices to define the flow. In the $2+1$ dimensional FRG analysis, the construction of the RG trajectory requires turning on, alongside $h$, a specific combination of other relevant deformations, e.g., $\Delta g_2$ and $\Delta g_3$.

\section{Thermal Effects}

The action \eqref{Smodel} preserves parity. At zero temperature, the effective potential of our construction is governed by CFT$_\text{IR}$, placing the system in a conformally invariant vacuum. This vacuum corresponds to the direct product of the critical GNY model and the critical vector model, both of which respect parity. Consequently, at $T=0$, the theory is manifestly invariant under~$\mathcal{P}$.

This situation changes at finite temperature. The effective potential is replaced by the free energy, which is computed by evaluating the path integral of the QFT on a cylinder whose radius is set by the inverse temperature. Consequently, the temperature -- or equivalently, the size of the cylinder -- sets the dominant energy scale, and naturally plays the role of the RG scale~$\mu$ at which the theory is probed.

Such a perspective aligns with the framework of critical phenomena, where $T$ controls the relevance of various deformations under coarse-graining (e.g., masses and Yukawa couplings in our model). This motivates the identification $\mu \sim T$, which we adopt in the following. A similar idea appears in studies of $c$-theorems \cite{Zamolodchikov:1986gt,Cardy:1988cwa,Komargodski:2011vj}, where one considers a QFT on a sphere \cite{Cardy:1988cwa,Klebanov:2011gs,Casini:2012ei,Casini:2017vbe}: the sphere’s radius acts as an RG scale, interpolating between UV and IR fixed points as the radius varies from zero to infinity.

In our case, increasing $T$ can be mapped to RG flow trajectories that run into the UV fixed point. At sufficiently high temperatures, the system becomes smeared over the excited states of $\text{CFT}_\text{UV}$, deformed by a small, relevant Yukawa coupling $h$. This coupling decreases with increasing $T$, and can be tuned to arbitrarily small values. As a result, the phase structure of the theory is governed entirely by the high-temperature behavior of $\text{CFT}_\text{UV}$. As shown in~\cite{Chai:2020zgq,Chai:2020onq,Hawashin:2024dpp}, CFT$_\text{UV}$ exhibits $\langle\chi\rangle_T \neq 0$ at finite temperatures. This spontaneously breaks $\mathcal{P}$-symmetry and generates a mass for the Dirac fermion. 

The mechanism of spontaneous parity breaking in our construction becomes particularly transparent within the $\epsilon$-expansion. Sufficiently close to 3+1 dimensions, the theory remains weakly coupled along the entire RG flow. This allows for a perturbative computation of the effective potential at any point along the RG trajectory. At finite temperatures, to leading order in the couplings, the effective potential is given by the classical potential supplemented by thermally induced masses~\cite{supplementary},
\begin{align}
    V_\text{eff}(\phi, \chi) &= \frac{\lambda_{\phi}}{8} (\phi^2)^2 + \frac{{\lambda_\chi}}{8} \chi^4 + \frac{\lambda_{\phi\chi}}{4} \chi^2 \phi^2 
    \nonumber \\
    &+ {1\over 2} m_\phi^2(T) \phi^2 + {1\over 2} m_\chi^2(T) \chi^2,
\end{align}
where 
\begin{align}
     m_\phi^2(T)&=\frac{T^2}{24} \left[\lambda_{\phi} (N_1 + 2) + \lambda_{\phi\chi} \right]~,  \nonumber \\
     m_\chi^2(T)&=\frac{T^2}{24} \left(3{\lambda_\chi} 
     + N_1 \lambda_{\phi\chi} + N_2 h^2\right) .
\end{align}
The thermal masses vanish at zero temperature, and the theory resides in the global minimum characterized by $\langle\chi\rangle = \langle\phi\rangle = 0$, with the couplings fixed at the critical values defined by CFT$_\text{IR}$. As the temperature increases, the thermal masses and couplings evolve. At sufficiently high temperatures, the Yukawa coupling $h$ becomes arbitrarily small, while the remaining couplings flow toward those of the conformal biconical vector model, with $\lambda^*_{\phi\chi} < 0$. Using \eqref{eq: betaeps}, one can verify that at these temperatures, $m_\chi^2(T) < 0$ for $N_1 \geq 18$. Consequently, $\langle\chi\rangle_T \neq 0$, resulting in a non-trivial mass for the Dirac field and indicating spontaneous parity breaking.

A similar observation can be inferred from the FRG approach. For $k < \pi T$, fermions decouple from the RG flow of the scalar sector due to the absence of zero fermionic Matsubara modes. This is reflected in~\eqref{eq: uflow}, where fermionic thermal factors vanish at such $k$ scales. Consider a QFT along the RG trajectory that we established to exist earlier, near CFT$_\text{UV}$, i.e., we consider small relevant deformations $\Delta g_2$, $\Delta g_3$, and $\Delta h$ away from CFT$_\text{UV}$. We have shown that at $T=0$, the IR phase of this QFT is governed by the parity-symmetric CFT$_\text{IR}$. At $ T\sim\Lambda$, fermions no longer affect the flow of the bosonic couplings, so their RG flow reduces to the flow of the biconical fixed point deformed along relevant directions in the space of bosonic couplings. The structure of the phase diagram near the biconical fixed point demonstrates that sufficiently small deformations lead to $\langle\chi\rangle_{T}\neq 0$ \cite{Hawashin:2024dpp}. Hence, there exists a critical intermediate temperature, which depends on the size of the deformation from CFT$_\text{UV}$, at which the theory transitions from a symmetric to a spontaneously parity-broken phase.

\section*{Acknowledgments} 

We thank Eliezer Rabinovici, Mireia Tolosa-Simeón, and Ritam Sinha for insightful discussion and correspondence. We also thank Zohar Komargodski for their valuable comments on the draft. MS acknowledges partial support from the BSF Grant No. 2022113 and NSF-BSF Grant No. 2022726. Additionally, MS and LY acknowledge partial support from Israel’s Council for Higher Education. BH and MMS are supported by the Mercator Research Center Ruhr under Project No.~Ko-2022-0012. MMS acknowledges funding from the Deutsche Forschungsgemeinschaft (DFG, German Research Foundation) under Project No.~277146847 (SFB 1238, project C02) and Project No.~452976698 (Heisenberg program).

\bibliography{bibliography}

\begin{thebibliography}{73}%
\makeatletter
\providecommand \@ifxundefined [1]{%
 \@ifx{#1\undefined}
}%
\providecommand \@ifnum [1]{%
 \ifnum #1\expandafter \@firstoftwo
 \else \expandafter \@secondoftwo
 \fi
}%
\providecommand \@ifx [1]{%
 \ifx #1\expandafter \@firstoftwo
 \else \expandafter \@secondoftwo
 \fi
}%
\providecommand \natexlab [1]{#1}%
\providecommand \enquote  [1]{``#1''}%
\providecommand \bibnamefont  [1]{#1}%
\providecommand \bibfnamefont [1]{#1}%
\providecommand \citenamefont [1]{#1}%
\providecommand \href@noop [0]{\@secondoftwo}%
\providecommand \href [0]{\begingroup \@sanitize@url \@href}%
\providecommand \@href[1]{\@@startlink{#1}\@@href}%
\providecommand \@@href[1]{\endgroup#1\@@endlink}%
\providecommand \@sanitize@url [0]{\catcode `\\12\catcode `\$12\catcode `\&12\catcode `\#12\catcode `\^12\catcode `\_12\catcode `\%12\relax}%
\providecommand \@@startlink[1]{}%
\providecommand \@@endlink[0]{}%
\providecommand \url  [0]{\begingroup\@sanitize@url \@url }%
\providecommand \@url [1]{\endgroup\@href {#1}{\urlprefix }}%
\providecommand \urlprefix  [0]{URL }%
\providecommand \Eprint [0]{\href }%
\providecommand \doibase [0]{https://doi.org/}%
\providecommand \selectlanguage [0]{\@gobble}%
\providecommand \bibinfo  [0]{\@secondoftwo}%
\providecommand \bibfield  [0]{\@secondoftwo}%
\providecommand \translation [1]{[#1]}%
\providecommand \BibitemOpen [0]{}%
\providecommand \bibitemStop [0]{}%
\providecommand \bibitemNoStop [0]{.\EOS\space}%
\providecommand \EOS [0]{\spacefactor3000\relax}%
\providecommand \BibitemShut  [1]{\csname bibitem#1\endcsname}%
\let\auto@bib@innerbib\@empty
%</preamble>
\bibitem [{\citenamefont {Landau}\ and\ \citenamefont {Lifshitz}(1980)}]{landau1980statistical}%
  \BibitemOpen
  \bibfield  {author} {\bibinfo {author} {\bibfnamefont {L.~D.}\ \bibnamefont {Landau}}\ and\ \bibinfo {author} {\bibfnamefont {E.~M.}\ \bibnamefont {Lifshitz}},\ }\href@noop {} {\emph {\bibinfo {title} {Statistical Physics, Part 1}}},\ \bibinfo {edition} {3rd}\ ed.,\ \bibinfo {series} {Course of Theoretical Physics}, Vol.~\bibinfo {volume} {5}\ (\bibinfo  {publisher} {Pergamon Press},\ \bibinfo {address} {Oxford},\ \bibinfo {year} {1980})\BibitemShut {NoStop}%
\bibitem [{\citenamefont {Pomeranchuk}(1950)}]{Pomeranchuk:1950}%
  \BibitemOpen
  \bibfield  {author} {\bibinfo {author} {\bibfnamefont {I.~Y.}\ \bibnamefont {Pomeranchuk}},\ }\bibfield  {title} {\bibinfo {title} {On the theory of liquid helium-3},\ }\href@noop {} {\bibfield  {journal} {\bibinfo  {journal} {Zh. Eksp. Teor. Fiz.}\ }\textbf {\bibinfo {volume} {20}},\ \bibinfo {pages} {919} (\bibinfo {year} {1950})},\ \bibinfo {note} {in Russian}\BibitemShut {NoStop}%
\bibitem [{\citenamefont {Weinberg}(1974)}]{weinberg1974gauge}%
  \BibitemOpen
  \bibfield  {author} {\bibinfo {author} {\bibfnamefont {S.}~\bibnamefont {Weinberg}},\ }\bibfield  {title} {\bibinfo {title} {Gauge and global symmetries at high temperature},\ }\href {https://doi.org/10.1103/PhysRevD.9.3357} {\bibfield  {journal} {\bibinfo  {journal} {Physical Review D}\ }\textbf {\bibinfo {volume} {9}},\ \bibinfo {pages} {3357} (\bibinfo {year} {1974})}\BibitemShut {NoStop}%
\bibitem [{\citenamefont {Han}\ \emph {et~al.}(2025)\citenamefont {Han}, \citenamefont {Huang}, \citenamefont {Komargodski}, \citenamefont {Lucas},\ and\ \citenamefont {Popov}}]{Han:2025eiw}%
  \BibitemOpen
  \bibfield  {author} {\bibinfo {author} {\bibfnamefont {Y.}~\bibnamefont {Han}}, \bibinfo {author} {\bibfnamefont {X.}~\bibnamefont {Huang}}, \bibinfo {author} {\bibfnamefont {Z.}~\bibnamefont {Komargodski}}, \bibinfo {author} {\bibfnamefont {A.}~\bibnamefont {Lucas}},\ and\ \bibinfo {author} {\bibfnamefont {F.~K.}\ \bibnamefont {Popov}},\ }\bibfield  {title} {\bibinfo {title} {{Entropic Order}},\ }\href@noop {} {\  (\bibinfo {year} {2025})},\ \Eprint {https://arxiv.org/abs/2503.22789} {arXiv:2503.22789 [cond-mat.stat-mech]} \BibitemShut {NoStop}%
\bibitem [{\citenamefont {Chai}\ \emph {et~al.}(2020{\natexlab{a}})\citenamefont {Chai}, \citenamefont {Chaudhuri}, \citenamefont {Choi}, \citenamefont {Komargodski}, \citenamefont {Rabinovici},\ and\ \citenamefont {Smolkin}}]{Chai:2020zgq}%
  \BibitemOpen
  \bibfield  {author} {\bibinfo {author} {\bibfnamefont {N.}~\bibnamefont {Chai}}, \bibinfo {author} {\bibfnamefont {S.}~\bibnamefont {Chaudhuri}}, \bibinfo {author} {\bibfnamefont {C.}~\bibnamefont {Choi}}, \bibinfo {author} {\bibfnamefont {Z.}~\bibnamefont {Komargodski}}, \bibinfo {author} {\bibfnamefont {E.}~\bibnamefont {Rabinovici}},\ and\ \bibinfo {author} {\bibfnamefont {M.}~\bibnamefont {Smolkin}},\ }\bibfield  {title} {\bibinfo {title} {{Thermal Order in Conformal Theories}},\ }\href {https://doi.org/10.1103/PhysRevD.102.065014} {\bibfield  {journal} {\bibinfo  {journal} {Phys. Rev. D}\ }\textbf {\bibinfo {volume} {102}},\ \bibinfo {pages} {065014} (\bibinfo {year} {2020}{\natexlab{a}})},\ \Eprint {https://arxiv.org/abs/2005.03676} {arXiv:2005.03676 [hep-th]} \BibitemShut {NoStop}%
\bibitem [{\citenamefont {Chai}\ \emph {et~al.}(2020{\natexlab{b}})\citenamefont {Chai}, \citenamefont {Chaudhuri}, \citenamefont {Choi}, \citenamefont {Komargodski}, \citenamefont {Rabinovici},\ and\ \citenamefont {Smolkin}}]{Chai:2020onq}%
  \BibitemOpen
  \bibfield  {author} {\bibinfo {author} {\bibfnamefont {N.}~\bibnamefont {Chai}}, \bibinfo {author} {\bibfnamefont {S.}~\bibnamefont {Chaudhuri}}, \bibinfo {author} {\bibfnamefont {C.}~\bibnamefont {Choi}}, \bibinfo {author} {\bibfnamefont {Z.}~\bibnamefont {Komargodski}}, \bibinfo {author} {\bibfnamefont {E.}~\bibnamefont {Rabinovici}},\ and\ \bibinfo {author} {\bibfnamefont {M.}~\bibnamefont {Smolkin}},\ }\bibfield  {title} {\bibinfo {title} {{Symmetry Breaking at All Temperatures}},\ }\href {https://doi.org/10.1103/PhysRevLett.125.131603} {\bibfield  {journal} {\bibinfo  {journal} {Phys. Rev. Lett.}\ }\textbf {\bibinfo {volume} {125}},\ \bibinfo {pages} {131603} (\bibinfo {year} {2020}{\natexlab{b}})}\BibitemShut {NoStop}%
\bibitem [{\citenamefont {Chai}\ \emph {et~al.}(2021)\citenamefont {Chai}, \citenamefont {Rabinovici}, \citenamefont {Sinha},\ and\ \citenamefont {Smolkin}}]{Chai:2020hnu}%
  \BibitemOpen
  \bibfield  {author} {\bibinfo {author} {\bibfnamefont {N.}~\bibnamefont {Chai}}, \bibinfo {author} {\bibfnamefont {E.}~\bibnamefont {Rabinovici}}, \bibinfo {author} {\bibfnamefont {R.}~\bibnamefont {Sinha}},\ and\ \bibinfo {author} {\bibfnamefont {M.}~\bibnamefont {Smolkin}},\ }\bibfield  {title} {\bibinfo {title} {{The bi-conical vector model at $1/N$}},\ }\href {https://doi.org/10.1007/JHEP05(2021)192} {\bibfield  {journal} {\bibinfo  {journal} {JHEP}\ }\textbf {\bibinfo {volume} {05}},\ \bibinfo {pages} {192}},\ \Eprint {https://arxiv.org/abs/2011.06003} {arXiv:2011.06003 [hep-th]} \BibitemShut {NoStop}%
\bibitem [{\citenamefont {Chai}\ \emph {et~al.}(2022{\natexlab{a}})\citenamefont {Chai}, \citenamefont {Dymarsky},\ and\ \citenamefont {Smolkin}}]{Chai:2021djc}%
  \BibitemOpen
  \bibfield  {author} {\bibinfo {author} {\bibfnamefont {N.}~\bibnamefont {Chai}}, \bibinfo {author} {\bibfnamefont {A.}~\bibnamefont {Dymarsky}},\ and\ \bibinfo {author} {\bibfnamefont {M.}~\bibnamefont {Smolkin}},\ }\bibfield  {title} {\bibinfo {title} {{Model of Persistent Breaking of Discrete Symmetry}},\ }\href {https://doi.org/10.1103/PhysRevLett.128.011601} {\bibfield  {journal} {\bibinfo  {journal} {Phys. Rev. Lett.}\ }\textbf {\bibinfo {volume} {128}},\ \bibinfo {pages} {011601} (\bibinfo {year} {2022}{\natexlab{a}})},\ \Eprint {https://arxiv.org/abs/2106.09723} {arXiv:2106.09723 [hep-th]} \BibitemShut {NoStop}%
\bibitem [{\citenamefont {Chai}\ \emph {et~al.}(2022{\natexlab{b}})\citenamefont {Chai}, \citenamefont {Dymarsky}, \citenamefont {Goykhman}, \citenamefont {Sinha},\ and\ \citenamefont {Smolkin}}]{Chai:2021tpt}%
  \BibitemOpen
  \bibfield  {author} {\bibinfo {author} {\bibfnamefont {N.}~\bibnamefont {Chai}}, \bibinfo {author} {\bibfnamefont {A.}~\bibnamefont {Dymarsky}}, \bibinfo {author} {\bibfnamefont {M.}~\bibnamefont {Goykhman}}, \bibinfo {author} {\bibfnamefont {R.}~\bibnamefont {Sinha}},\ and\ \bibinfo {author} {\bibfnamefont {M.}~\bibnamefont {Smolkin}},\ }\bibfield  {title} {\bibinfo {title} {{A model of persistent breaking of continuous symmetry}},\ }\href {https://doi.org/10.21468/SciPostPhys.12.6.181} {\bibfield  {journal} {\bibinfo  {journal} {SciPost Phys.}\ }\textbf {\bibinfo {volume} {12}},\ \bibinfo {pages} {181} (\bibinfo {year} {2022}{\natexlab{b}})},\ \Eprint {https://arxiv.org/abs/2111.02474} {arXiv:2111.02474 [hep-th]} \BibitemShut {NoStop}%
\bibitem [{\citenamefont {Liendo}\ \emph {et~al.}(2023)\citenamefont {Liendo}, \citenamefont {Rong},\ and\ \citenamefont {Zhang}}]{Liendo:2022bmv}%
  \BibitemOpen
  \bibfield  {author} {\bibinfo {author} {\bibfnamefont {P.}~\bibnamefont {Liendo}}, \bibinfo {author} {\bibfnamefont {J.}~\bibnamefont {Rong}},\ and\ \bibinfo {author} {\bibfnamefont {H.}~\bibnamefont {Zhang}},\ }\bibfield  {title} {\bibinfo {title} {{Spontaneous breaking of finite group symmetries at all temperatures}},\ }\href {https://doi.org/10.21468/SciPostPhys.14.6.168} {\bibfield  {journal} {\bibinfo  {journal} {SciPost Phys.}\ }\textbf {\bibinfo {volume} {14}},\ \bibinfo {pages} {168} (\bibinfo {year} {2023})},\ \Eprint {https://arxiv.org/abs/2205.13964} {arXiv:2205.13964 [hep-th]} \BibitemShut {NoStop}%
\bibitem [{\citenamefont {Hawashin}\ \emph {et~al.}(2025)\citenamefont {Hawashin}, \citenamefont {Rong},\ and\ \citenamefont {Scherer}}]{Hawashin:2024dpp}%
  \BibitemOpen
  \bibfield  {author} {\bibinfo {author} {\bibfnamefont {B.}~\bibnamefont {Hawashin}}, \bibinfo {author} {\bibfnamefont {J.}~\bibnamefont {Rong}},\ and\ \bibinfo {author} {\bibfnamefont {M.~M.}\ \bibnamefont {Scherer}},\ }\bibfield  {title} {\bibinfo {title} {Ultraviolet-complete local field theory of persistent symmetry breaking in $2+1$ dimensions},\ }\href {https://doi.org/10.1103/PhysRevLett.134.041602} {\bibfield  {journal} {\bibinfo  {journal} {Phys. Rev. Lett.}\ }\textbf {\bibinfo {volume} {134}},\ \bibinfo {pages} {041602} (\bibinfo {year} {2025})}\BibitemShut {NoStop}%
\bibitem [{\citenamefont {Komargodski}\ and\ \citenamefont {Popov}(2024)}]{Komargodski:2024zmt}%
  \BibitemOpen
  \bibfield  {author} {\bibinfo {author} {\bibfnamefont {Z.}~\bibnamefont {Komargodski}}\ and\ \bibinfo {author} {\bibfnamefont {F.~K.}\ \bibnamefont {Popov}},\ }\bibfield  {title} {\bibinfo {title} {{Temperature-Resistant Order in 2+1 Dimensions}},\ }\href@noop {} {\  (\bibinfo {year} {2024})},\ \Eprint {https://arxiv.org/abs/2412.09459} {arXiv:2412.09459 [hep-th]} \BibitemShut {NoStop}%
\bibitem [{\citenamefont {Hogervorst}\ \emph {et~al.}(2016)\citenamefont {Hogervorst}, \citenamefont {Rychkov},\ and\ \citenamefont {van Rees}}]{Hogervorst:2015akt}%
  \BibitemOpen
  \bibfield  {author} {\bibinfo {author} {\bibfnamefont {M.}~\bibnamefont {Hogervorst}}, \bibinfo {author} {\bibfnamefont {S.}~\bibnamefont {Rychkov}},\ and\ \bibinfo {author} {\bibfnamefont {B.~C.}\ \bibnamefont {van Rees}},\ }\bibfield  {title} {\bibinfo {title} {{Unitarity violation at the Wilson-Fisher fixed point in 4-$\epsilon$ dimensions}},\ }\href {https://doi.org/10.1103/PhysRevD.93.125025} {\bibfield  {journal} {\bibinfo  {journal} {Phys. Rev. D}\ }\textbf {\bibinfo {volume} {93}},\ \bibinfo {pages} {125025} (\bibinfo {year} {2016})},\ \Eprint {https://arxiv.org/abs/1512.00013} {arXiv:1512.00013 [hep-th]} \BibitemShut {NoStop}%
\bibitem [{\citenamefont {Belavin}\ and\ \citenamefont {Migdal}(1974)}]{Belavin:1974gu}%
  \BibitemOpen
  \bibfield  {author} {\bibinfo {author} {\bibfnamefont {A.~A.}\ \bibnamefont {Belavin}}\ and\ \bibinfo {author} {\bibfnamefont {A.~A.}\ \bibnamefont {Migdal}},\ }\bibfield  {title} {\bibinfo {title} {{Calculation of anomalous dimensions in non-abelian gauge field theories}},\ }\href@noop {} {\bibfield  {journal} {\bibinfo  {journal} {Pisma Zh. Eksp. Teor. Fiz.}\ }\textbf {\bibinfo {volume} {19}},\ \bibinfo {pages} {317} (\bibinfo {year} {1974})}\BibitemShut {NoStop}%
\bibitem [{\citenamefont {Caswell}(1974)}]{Caswell:1974gg}%
  \BibitemOpen
  \bibfield  {author} {\bibinfo {author} {\bibfnamefont {W.~E.}\ \bibnamefont {Caswell}},\ }\bibfield  {title} {\bibinfo {title} {Asymptotic behavior of non-abelian gauge theories to two-loop order},\ }\href {https://doi.org/10.1103/PhysRevLett.33.244} {\bibfield  {journal} {\bibinfo  {journal} {Phys. Rev. Lett.}\ }\textbf {\bibinfo {volume} {33}},\ \bibinfo {pages} {244} (\bibinfo {year} {1974})}\BibitemShut {NoStop}%
\bibitem [{\citenamefont {Banks}\ and\ \citenamefont {Zaks}(1982)}]{Banks:1981nn}%
  \BibitemOpen
  \bibfield  {author} {\bibinfo {author} {\bibfnamefont {T.}~\bibnamefont {Banks}}\ and\ \bibinfo {author} {\bibfnamefont {A.}~\bibnamefont {Zaks}},\ }\bibfield  {title} {\bibinfo {title} {{On the Phase Structure of Vector-Like Gauge Theories with Massless Fermions}},\ }\href {https://doi.org/10.1016/0550-3213(82)90035-9} {\bibfield  {journal} {\bibinfo  {journal} {Nucl. Phys. B}\ }\textbf {\bibinfo {volume} {196}},\ \bibinfo {pages} {189} (\bibinfo {year} {1982})}\BibitemShut {NoStop}%
%%CITATION = NUPHA,B196,189;%%
\bibitem [{\citenamefont {Chaudhuri}\ \emph {et~al.}(2021)\citenamefont {Chaudhuri}, \citenamefont {Choi},\ and\ \citenamefont {Rabinovici}}]{Chaudhuri:2020xxb}%
  \BibitemOpen
  \bibfield  {author} {\bibinfo {author} {\bibfnamefont {S.}~\bibnamefont {Chaudhuri}}, \bibinfo {author} {\bibfnamefont {C.}~\bibnamefont {Choi}},\ and\ \bibinfo {author} {\bibfnamefont {E.}~\bibnamefont {Rabinovici}},\ }\bibfield  {title} {\bibinfo {title} {{Thermal order in large N conformal gauge theories}},\ }\href {https://doi.org/10.1007/JHEP04(2021)203} {\bibfield  {journal} {\bibinfo  {journal} {JHEP}\ }\textbf {\bibinfo {volume} {04}},\ \bibinfo {pages} {203}},\ \Eprint {https://arxiv.org/abs/2011.13981} {arXiv:2011.13981 [hep-th]} \BibitemShut {NoStop}%
\bibitem [{\citenamefont {Bajc}\ \emph {et~al.}(2021)\citenamefont {Bajc}, \citenamefont {Lugo},\ and\ \citenamefont {Sannino}}]{Bajc:2020gpa}%
  \BibitemOpen
  \bibfield  {author} {\bibinfo {author} {\bibfnamefont {B.}~\bibnamefont {Bajc}}, \bibinfo {author} {\bibfnamefont {A.}~\bibnamefont {Lugo}},\ and\ \bibinfo {author} {\bibfnamefont {F.}~\bibnamefont {Sannino}},\ }\bibfield  {title} {\bibinfo {title} {{Asymptotically free and safe fate of symmetry nonrestoration}},\ }\href {https://doi.org/10.1103/PhysRevD.103.096014} {\bibfield  {journal} {\bibinfo  {journal} {Phys. Rev. D}\ }\textbf {\bibinfo {volume} {103}},\ \bibinfo {pages} {096014} (\bibinfo {year} {2021})},\ \Eprint {https://arxiv.org/abs/2012.08428} {arXiv:2012.08428 [hep-th]} \BibitemShut {NoStop}%
\bibitem [{\citenamefont {Chaudhuri}\ and\ \citenamefont {Rabinovici}(2021)}]{Chaudhuri:2021dsq}%
  \BibitemOpen
  \bibfield  {author} {\bibinfo {author} {\bibfnamefont {S.}~\bibnamefont {Chaudhuri}}\ and\ \bibinfo {author} {\bibfnamefont {E.}~\bibnamefont {Rabinovici}},\ }\bibfield  {title} {\bibinfo {title} {{Symmetry breaking at high temperatures in large N gauge theories}},\ }\href {https://doi.org/10.1007/JHEP08(2021)148} {\bibfield  {journal} {\bibinfo  {journal} {JHEP}\ }\textbf {\bibinfo {volume} {08}},\ \bibinfo {pages} {148}},\ \Eprint {https://arxiv.org/abs/2106.11323} {arXiv:2106.11323 [hep-th]} \BibitemShut {NoStop}%
\bibitem [{\citenamefont {Moshe}\ and\ \citenamefont {Zinn-Justin}(2003)}]{Moshe:2003xn}%
  \BibitemOpen
  \bibfield  {author} {\bibinfo {author} {\bibfnamefont {M.}~\bibnamefont {Moshe}}\ and\ \bibinfo {author} {\bibfnamefont {J.}~\bibnamefont {Zinn-Justin}},\ }\bibfield  {title} {\bibinfo {title} {{Quantum field theory in the large N limit: A Review}},\ }\href {https://doi.org/10.1016/S0370-1573(03)00263-1} {\bibfield  {journal} {\bibinfo  {journal} {Phys. Rept.}\ }\textbf {\bibinfo {volume} {385}},\ \bibinfo {pages} {69} (\bibinfo {year} {2003})},\ \Eprint {https://arxiv.org/abs/hep-th/0306133} {arXiv:hep-th/0306133} \BibitemShut {NoStop}%
\bibitem [{\citenamefont {Nelson}\ \emph {et~al.}(1974)\citenamefont {Nelson}, \citenamefont {Kosterlitz},\ and\ \citenamefont {Fisher}}]{Fisher_mcl:1974}%
  \BibitemOpen
  \bibfield  {author} {\bibinfo {author} {\bibfnamefont {D.~R.}\ \bibnamefont {Nelson}}, \bibinfo {author} {\bibfnamefont {J.~M.}\ \bibnamefont {Kosterlitz}},\ and\ \bibinfo {author} {\bibfnamefont {M.~E.}\ \bibnamefont {Fisher}},\ }\bibfield  {title} {\bibinfo {title} {Renormalization-group analysis of bicritical and tetracritical points},\ }\href {https://doi.org/10.1103/PhysRevLett.33.813} {\bibfield  {journal} {\bibinfo  {journal} {Phys. Rev. Lett.}\ }\textbf {\bibinfo {volume} {33}},\ \bibinfo {pages} {813} (\bibinfo {year} {1974})}\BibitemShut {NoStop}%
\bibitem [{\citenamefont {Calabrese}\ \emph {et~al.}(2003)\citenamefont {Calabrese}, \citenamefont {Pelissetto},\ and\ \citenamefont {Vicari}}]{Vicari:2003}%
  \BibitemOpen
  \bibfield  {author} {\bibinfo {author} {\bibfnamefont {P.}~\bibnamefont {Calabrese}}, \bibinfo {author} {\bibfnamefont {A.}~\bibnamefont {Pelissetto}},\ and\ \bibinfo {author} {\bibfnamefont {E.}~\bibnamefont {Vicari}},\ }\bibfield  {title} {\bibinfo {title} {Multicritical phenomena in $\mathrm{O}{(n}_{1})\ensuremath{\bigoplus}\mathrm{O}{(n}_{2})$-symmetric theories},\ }\href {https://doi.org/10.1103/PhysRevB.67.054505} {\bibfield  {journal} {\bibinfo  {journal} {Phys. Rev. B}\ }\textbf {\bibinfo {volume} {67}},\ \bibinfo {pages} {054505} (\bibinfo {year} {2003})}\BibitemShut {NoStop}%
\bibitem [{\citenamefont {Eichhorn}\ \emph {et~al.}(2013)\citenamefont {Eichhorn}, \citenamefont {Mesterh\'azy},\ and\ \citenamefont {Scherer}}]{Eichhorn_mcl2013}%
  \BibitemOpen
  \bibfield  {author} {\bibinfo {author} {\bibfnamefont {A.}~\bibnamefont {Eichhorn}}, \bibinfo {author} {\bibfnamefont {D.}~\bibnamefont {Mesterh\'azy}},\ and\ \bibinfo {author} {\bibfnamefont {M.~M.}\ \bibnamefont {Scherer}},\ }\bibfield  {title} {\bibinfo {title} {Multicritical behavior in models with two competing order parameters},\ }\href {https://doi.org/10.1103/PhysRevE.88.042141} {\bibfield  {journal} {\bibinfo  {journal} {Phys. Rev. E}\ }\textbf {\bibinfo {volume} {88}},\ \bibinfo {pages} {042141} (\bibinfo {year} {2013})}\BibitemShut {NoStop}%
\bibitem [{\citenamefont {Meade}\ and\ \citenamefont {Ramani}(2019)}]{Meade:2018saz}%
  \BibitemOpen
  \bibfield  {author} {\bibinfo {author} {\bibfnamefont {P.}~\bibnamefont {Meade}}\ and\ \bibinfo {author} {\bibfnamefont {H.}~\bibnamefont {Ramani}},\ }\bibfield  {title} {\bibinfo {title} {{Unrestored Electroweak Symmetry}},\ }\href {https://doi.org/10.1103/PhysRevLett.122.041802} {\bibfield  {journal} {\bibinfo  {journal} {Phys. Rev. Lett.}\ }\textbf {\bibinfo {volume} {122}},\ \bibinfo {pages} {041802} (\bibinfo {year} {2019})},\ \Eprint {https://arxiv.org/abs/1807.07578} {arXiv:1807.07578 [hep-ph]} \BibitemShut {NoStop}%
\bibitem [{\citenamefont {Bai}\ \emph {et~al.}(2021)\citenamefont {Bai}, \citenamefont {Lee}, \citenamefont {Son},\ and\ \citenamefont {Ye}}]{Bai:2021hfb}%
  \BibitemOpen
  \bibfield  {author} {\bibinfo {author} {\bibfnamefont {Y.}~\bibnamefont {Bai}}, \bibinfo {author} {\bibfnamefont {S.~J.}\ \bibnamefont {Lee}}, \bibinfo {author} {\bibfnamefont {M.}~\bibnamefont {Son}},\ and\ \bibinfo {author} {\bibfnamefont {F.}~\bibnamefont {Ye}},\ }\bibfield  {title} {\bibinfo {title} {{Global electroweak symmetric vacuum}},\ }\href {https://doi.org/10.1007/JHEP07(2021)225} {\bibfield  {journal} {\bibinfo  {journal} {JHEP}\ }\textbf {\bibinfo {volume} {07}},\ \bibinfo {pages} {225}},\ \Eprint {https://arxiv.org/abs/2103.09819} {arXiv:2103.09819 [hep-ph]} \BibitemShut {NoStop}%
\bibitem [{\citenamefont {Ramazanov}\ \emph {et~al.}(2021)\citenamefont {Ramazanov}, \citenamefont {Urban},\ and\ \citenamefont {Vikman}}]{Ramazanov:2020ajq}%
  \BibitemOpen
  \bibfield  {author} {\bibinfo {author} {\bibfnamefont {S.}~\bibnamefont {Ramazanov}}, \bibinfo {author} {\bibfnamefont {F.~R.}\ \bibnamefont {Urban}},\ and\ \bibinfo {author} {\bibfnamefont {A.}~\bibnamefont {Vikman}},\ }\bibfield  {title} {\bibinfo {title} {{Observing primordial magnetic fields through Dark Matter}},\ }\href {https://doi.org/10.1088/1475-7516/2021/02/011} {\bibfield  {journal} {\bibinfo  {journal} {JCAP}\ }\textbf {\bibinfo {volume} {02}},\ \bibinfo {pages} {011}},\ \Eprint {https://arxiv.org/abs/2010.03383} {arXiv:2010.03383 [astro-ph.CO]} \BibitemShut {NoStop}%
\bibitem [{\citenamefont {Ramazanov}\ \emph {et~al.}(2022)\citenamefont {Ramazanov}, \citenamefont {Babichev}, \citenamefont {Gorbunov},\ and\ \citenamefont {Vikman}}]{Ramazanov:2021eya}%
  \BibitemOpen
  \bibfield  {author} {\bibinfo {author} {\bibfnamefont {S.}~\bibnamefont {Ramazanov}}, \bibinfo {author} {\bibfnamefont {E.}~\bibnamefont {Babichev}}, \bibinfo {author} {\bibfnamefont {D.}~\bibnamefont {Gorbunov}},\ and\ \bibinfo {author} {\bibfnamefont {A.}~\bibnamefont {Vikman}},\ }\bibfield  {title} {\bibinfo {title} {{Beyond freeze-in: Dark matter via inverse phase transition and gravitational wave signal}},\ }\href {https://doi.org/10.1103/PhysRevD.105.063530} {\bibfield  {journal} {\bibinfo  {journal} {Phys. Rev. D}\ }\textbf {\bibinfo {volume} {105}},\ \bibinfo {pages} {063530} (\bibinfo {year} {2022})},\ \Eprint {https://arxiv.org/abs/2104.13722} {arXiv:2104.13722 [hep-ph]} \BibitemShut {NoStop}%
\bibitem [{\citenamefont {Mermin}\ and\ \citenamefont {Wagner}(1966)}]{PhysRevLett.17.1133}%
  \BibitemOpen
  \bibfield  {author} {\bibinfo {author} {\bibfnamefont {N.~D.}\ \bibnamefont {Mermin}}\ and\ \bibinfo {author} {\bibfnamefont {H.}~\bibnamefont {Wagner}},\ }\bibfield  {title} {\bibinfo {title} {Absence of ferromagnetism or antiferromagnetism in one- or two-dimensional isotropic heisenberg models},\ }\href {https://doi.org/10.1103/PhysRevLett.17.1133} {\bibfield  {journal} {\bibinfo  {journal} {Phys. Rev. Lett.}\ }\textbf {\bibinfo {volume} {17}},\ \bibinfo {pages} {1133} (\bibinfo {year} {1966})}\BibitemShut {NoStop}%
\bibitem [{\citenamefont {Hohenberg}(1967)}]{PhysRev.158.383}%
  \BibitemOpen
  \bibfield  {author} {\bibinfo {author} {\bibfnamefont {P.~C.}\ \bibnamefont {Hohenberg}},\ }\bibfield  {title} {\bibinfo {title} {Existence of long-range order in one and two dimensions},\ }\href {https://doi.org/10.1103/PhysRev.158.383} {\bibfield  {journal} {\bibinfo  {journal} {Phys. Rev.}\ }\textbf {\bibinfo {volume} {158}},\ \bibinfo {pages} {383} (\bibinfo {year} {1967})}\BibitemShut {NoStop}%
\bibitem [{\citenamefont {Coleman}(1973)}]{Coleman:1973ci}%
  \BibitemOpen
  \bibfield  {author} {\bibinfo {author} {\bibfnamefont {S.~R.}\ \bibnamefont {Coleman}},\ }\bibfield  {title} {\bibinfo {title} {{There are no Goldstone bosons in two-dimensions}},\ }\href {https://doi.org/10.1007/BF01646487} {\bibfield  {journal} {\bibinfo  {journal} {Commun. Math. Phys.}\ }\textbf {\bibinfo {volume} {31}},\ \bibinfo {pages} {259} (\bibinfo {year} {1973})}\BibitemShut {NoStop}%
\bibitem [{\citenamefont {Buchel}(2021{\natexlab{a}})}]{Buchel:2020thm}%
  \BibitemOpen
  \bibfield  {author} {\bibinfo {author} {\bibfnamefont {A.}~\bibnamefont {Buchel}},\ }\bibfield  {title} {\bibinfo {title} {{Thermal order in holographic CFTs and no-hair theorem violation in black branes}},\ }\href {https://doi.org/10.1016/j.nuclphysb.2021.115425} {\bibfield  {journal} {\bibinfo  {journal} {Nucl. Phys. B}\ }\textbf {\bibinfo {volume} {967}},\ \bibinfo {pages} {115425} (\bibinfo {year} {2021}{\natexlab{a}})},\ \Eprint {https://arxiv.org/abs/2005.07833} {arXiv:2005.07833 [hep-th]} \BibitemShut {NoStop}%
\bibitem [{\citenamefont {Buchel}(2021{\natexlab{b}})}]{Buchel:2020xdk}%
  \BibitemOpen
  \bibfield  {author} {\bibinfo {author} {\bibfnamefont {A.}~\bibnamefont {Buchel}},\ }\bibfield  {title} {\bibinfo {title} {{Holographic conformal order in supergravity}},\ }\href {https://doi.org/10.1016/j.physletb.2021.136111} {\bibfield  {journal} {\bibinfo  {journal} {Phys. Lett. B}\ }\textbf {\bibinfo {volume} {814}},\ \bibinfo {pages} {136111} (\bibinfo {year} {2021}{\natexlab{b}})},\ \Eprint {https://arxiv.org/abs/2007.09420} {arXiv:2007.09420 [hep-th]} \BibitemShut {NoStop}%
\bibitem [{\citenamefont {Buchel}(2021{\natexlab{c}})}]{Buchel:2020jfs}%
  \BibitemOpen
  \bibfield  {author} {\bibinfo {author} {\bibfnamefont {A.}~\bibnamefont {Buchel}},\ }\bibfield  {title} {\bibinfo {title} {{Fate of the conformal order}},\ }\href {https://doi.org/10.1103/PhysRevD.103.026008} {\bibfield  {journal} {\bibinfo  {journal} {Phys. Rev. D}\ }\textbf {\bibinfo {volume} {103}},\ \bibinfo {pages} {026008} (\bibinfo {year} {2021}{\natexlab{c}})},\ \Eprint {https://arxiv.org/abs/2011.11509} {arXiv:2011.11509 [hep-th]} \BibitemShut {NoStop}%
\bibitem [{\citenamefont {Buchel}(2021{\natexlab{d}})}]{Buchel:2021ead}%
  \BibitemOpen
  \bibfield  {author} {\bibinfo {author} {\bibfnamefont {A.}~\bibnamefont {Buchel}},\ }\bibfield  {title} {\bibinfo {title} {{Compactified holographic conformal order}},\ }\href {https://doi.org/10.1016/j.nuclphysb.2021.115605} {\bibfield  {journal} {\bibinfo  {journal} {Nucl. Phys. B}\ }\textbf {\bibinfo {volume} {973}},\ \bibinfo {pages} {115605} (\bibinfo {year} {2021}{\natexlab{d}})},\ \Eprint {https://arxiv.org/abs/2107.05086} {arXiv:2107.05086 [hep-th]} \BibitemShut {NoStop}%
\bibitem [{\citenamefont {Buchel}(2022)}]{Buchel:2022zxl}%
  \BibitemOpen
  \bibfield  {author} {\bibinfo {author} {\bibfnamefont {A.}~\bibnamefont {Buchel}},\ }\bibfield  {title} {\bibinfo {title} {{The quest for a conifold conformal order}},\ }\href {https://doi.org/10.1007/JHEP08(2022)080} {\bibfield  {journal} {\bibinfo  {journal} {JHEP}\ }\textbf {\bibinfo {volume} {08}},\ \bibinfo {pages} {080}},\ \Eprint {https://arxiv.org/abs/2205.00612} {arXiv:2205.00612 [hep-th]} \BibitemShut {NoStop}%
\bibitem [{\citenamefont {Buchel}(2024)}]{Buchel:2023zpe}%
  \BibitemOpen
  \bibfield  {author} {\bibinfo {author} {\bibfnamefont {A.}~\bibnamefont {Buchel}},\ }\bibfield  {title} {\bibinfo {title} {{Holographic conformal order with higher derivatives}},\ }\href {https://doi.org/10.1016/j.nuclphysb.2024.116578} {\bibfield  {journal} {\bibinfo  {journal} {Nucl. Phys. B}\ }\textbf {\bibinfo {volume} {1004}},\ \bibinfo {pages} {116578} (\bibinfo {year} {2024})},\ \Eprint {https://arxiv.org/abs/2312.15764} {arXiv:2312.15764 [hep-th]} \BibitemShut {NoStop}%
\bibitem [{\citenamefont {Buchel}(2025)}]{Buchel:2025cve}%
  \BibitemOpen
  \bibfield  {author} {\bibinfo {author} {\bibfnamefont {A.}~\bibnamefont {Buchel}},\ }\bibfield  {title} {\bibinfo {title} {{The ordered phase of charged N=4 SYM plasma}},\ }\href@noop {} {\  (\bibinfo {year} {2025})},\ \Eprint {https://arxiv.org/abs/2501.01856} {arXiv:2501.01856 [hep-th]} \BibitemShut {NoStop}%
\bibitem [{Note1()}]{Note1}%
  \BibitemOpen
  \bibinfo {note} {For brevity, we omit the explicit flavor indices of $\phi $ and $\psi $; summation over flavor contractions is implied throughout.}\BibitemShut {Stop}%
\bibitem [{\citenamefont {Zinn-Justin}(1991)}]{Zinn-Justin:1991ksq}%
  \BibitemOpen
  \bibfield  {author} {\bibinfo {author} {\bibfnamefont {J.}~\bibnamefont {Zinn-Justin}},\ }\bibfield  {title} {\bibinfo {title} {{Four fermion interaction near four-dimensions}},\ }\href {https://doi.org/10.1016/0550-3213(91)90043-W} {\bibfield  {journal} {\bibinfo  {journal} {Nucl. Phys. B}\ }\textbf {\bibinfo {volume} {367}},\ \bibinfo {pages} {105} (\bibinfo {year} {1991})}\BibitemShut {NoStop}%
\bibitem [{\citenamefont {Hands}\ \emph {et~al.}(1993)\citenamefont {Hands}, \citenamefont {Kocic},\ and\ \citenamefont {Kogut}}]{Hands:1992be}%
  \BibitemOpen
  \bibfield  {author} {\bibinfo {author} {\bibfnamefont {S.}~\bibnamefont {Hands}}, \bibinfo {author} {\bibfnamefont {A.}~\bibnamefont {Kocic}},\ and\ \bibinfo {author} {\bibfnamefont {J.~B.}\ \bibnamefont {Kogut}},\ }\bibfield  {title} {\bibinfo {title} {{Four Fermi theories in fewer than four-dimensions}},\ }\href {https://doi.org/10.1006/aphy.1993.1039} {\bibfield  {journal} {\bibinfo  {journal} {Annals Phys.}\ }\textbf {\bibinfo {volume} {224}},\ \bibinfo {pages} {29} (\bibinfo {year} {1993})},\ \Eprint {https://arxiv.org/abs/hep-lat/9208022} {arXiv:hep-lat/9208022} \BibitemShut {NoStop}%
\bibitem [{\citenamefont {Karkkainen}\ \emph {et~al.}(1994)\citenamefont {Karkkainen}, \citenamefont {Lacaze}, \citenamefont {Lacock},\ and\ \citenamefont {Petersson}}]{Karkkainen:1993ef}%
  \BibitemOpen
  \bibfield  {author} {\bibinfo {author} {\bibfnamefont {L.}~\bibnamefont {Karkkainen}}, \bibinfo {author} {\bibfnamefont {R.}~\bibnamefont {Lacaze}}, \bibinfo {author} {\bibfnamefont {P.}~\bibnamefont {Lacock}},\ and\ \bibinfo {author} {\bibfnamefont {B.}~\bibnamefont {Petersson}},\ }\bibfield  {title} {\bibinfo {title} {{Critical behavior of the three-dimensional Gross-Neveu and Higgs-Yukawa models}},\ }\href {https://doi.org/10.1016/0550-3213(94)90309-3} {\bibfield  {journal} {\bibinfo  {journal} {Nucl. Phys. B}\ }\textbf {\bibinfo {volume} {415}},\ \bibinfo {pages} {781} (\bibinfo {year} {1994})},\ \bibinfo {note} {[Erratum: Nucl.Phys.B 438, 650--650 (1995)]},\ \Eprint {https://arxiv.org/abs/hep-lat/9310020} {arXiv:hep-lat/9310020} \BibitemShut {NoStop}%
\bibitem [{\citenamefont {GRACEY}(1994{\natexlab{a}})}]{doi:10.1142/S0217751X94000285}%
  \BibitemOpen
  \bibfield  {author} {\bibinfo {author} {\bibfnamefont {J.}~\bibnamefont {GRACEY}},\ }\bibfield  {title} {\bibinfo {title} {Computation of $\beta'$(gc) at o(1/n2) in the o(n) gross-neveu model in arbitrary dimensions},\ }\href {https://doi.org/10.1142/S0217751X94000285} {\bibfield  {journal} {\bibinfo  {journal} {International Journal of Modern Physics A}\ }\textbf {\bibinfo {volume} {09}},\ \bibinfo {pages} {567} (\bibinfo {year} {1994}{\natexlab{a}})},\ \Eprint {https://arxiv.org/abs/hep-th/9306106} {arXiv:hep-th/9306106} \BibitemShut {NoStop}%
\bibitem [{\citenamefont {GRACEY}(1994{\natexlab{b}})}]{doi:10.1142/S0217751X94000340}%
  \BibitemOpen
  \bibfield  {author} {\bibinfo {author} {\bibfnamefont {J.}~\bibnamefont {GRACEY}},\ }\bibfield  {title} {\bibinfo {title} {Computation of critical exponent $\eta$ at o(1/n3) in the four-fermi model in arbitrary dimensions},\ }\href {https://doi.org/10.1142/S0217751X94000340} {\bibfield  {journal} {\bibinfo  {journal} {International Journal of Modern Physics A}\ }\textbf {\bibinfo {volume} {09}},\ \bibinfo {pages} {727} (\bibinfo {year} {1994}{\natexlab{b}})},\ \Eprint {https://arxiv.org/abs/hep-th/9306107} {arXiv:hep-th/9306107} \BibitemShut {NoStop}%
\bibitem [{\citenamefont {Hasenbusch}(2010)}]{Hasenbusch:2010hkh}%
  \BibitemOpen
  \bibfield  {author} {\bibinfo {author} {\bibfnamefont {M.}~\bibnamefont {Hasenbusch}},\ }\bibfield  {title} {\bibinfo {title} {{Finite size scaling study of lattice models in the three-dimensional Ising universality class}},\ }\href {https://doi.org/10.1103/PhysRevB.82.174433} {\bibfield  {journal} {\bibinfo  {journal} {Phys. Rev. B}\ }\textbf {\bibinfo {volume} {82}},\ \bibinfo {pages} {174433} (\bibinfo {year} {2010})},\ \Eprint {https://arxiv.org/abs/1004.4486} {arXiv:1004.4486 [cond-mat.stat-mech]} \BibitemShut {NoStop}%
\bibitem [{\citenamefont {Kos}\ \emph {et~al.}(2014)\citenamefont {Kos}, \citenamefont {Poland},\ and\ \citenamefont {Simmons-Duffin}}]{Kos:2013tga}%
  \BibitemOpen
  \bibfield  {author} {\bibinfo {author} {\bibfnamefont {F.}~\bibnamefont {Kos}}, \bibinfo {author} {\bibfnamefont {D.}~\bibnamefont {Poland}},\ and\ \bibinfo {author} {\bibfnamefont {D.}~\bibnamefont {Simmons-Duffin}},\ }\bibfield  {title} {\bibinfo {title} {{Bootstrapping the $O(N)$ vector models}},\ }\href {https://doi.org/10.1007/JHEP06(2014)091} {\bibfield  {journal} {\bibinfo  {journal} {JHEP}\ }\textbf {\bibinfo {volume} {06}},\ \bibinfo {pages} {091}},\ \Eprint {https://arxiv.org/abs/1307.6856} {arXiv:1307.6856 [hep-th]} \BibitemShut {NoStop}%
\bibitem [{\citenamefont {Janssen}\ and\ \citenamefont {Herbut}(2014{\natexlab{a}})}]{Janssen:2014gea}%
  \BibitemOpen
  \bibfield  {author} {\bibinfo {author} {\bibfnamefont {L.}~\bibnamefont {Janssen}}\ and\ \bibinfo {author} {\bibfnamefont {I.~F.}\ \bibnamefont {Herbut}},\ }\bibfield  {title} {\bibinfo {title} {{Antiferromagnetic critical point on graphene's honeycomb lattice: A functional renormalization group approach}},\ }\href {https://doi.org/10.1103/PhysRevB.89.205403} {\bibfield  {journal} {\bibinfo  {journal} {Phys. Rev. B}\ }\textbf {\bibinfo {volume} {89}},\ \bibinfo {pages} {205403} (\bibinfo {year} {2014}{\natexlab{a}})},\ \bibinfo {note} {[Addendum: Phys.Rev.B 102, 199902 (2020)]},\ \Eprint {https://arxiv.org/abs/1402.6277} {arXiv:1402.6277 [cond-mat.str-el]} \BibitemShut {NoStop}%
\bibitem [{\citenamefont {Knorr}(2016)}]{PhysRevB.94.245102}%
  \BibitemOpen
  \bibfield  {author} {\bibinfo {author} {\bibfnamefont {B.}~\bibnamefont {Knorr}},\ }\bibfield  {title} {\bibinfo {title} {Ising and gross-neveu model in next-to-leading order},\ }\href {https://doi.org/10.1103/PhysRevB.94.245102} {\bibfield  {journal} {\bibinfo  {journal} {Phys. Rev. B}\ }\textbf {\bibinfo {volume} {94}},\ \bibinfo {pages} {245102} (\bibinfo {year} {2016})}\BibitemShut {NoStop}%
\bibitem [{\citenamefont {Ihrig}\ \emph {et~al.}(2018)\citenamefont {Ihrig}, \citenamefont {Mihaila},\ and\ \citenamefont {Scherer}}]{PhysRevB.98.125109}%
  \BibitemOpen
  \bibfield  {author} {\bibinfo {author} {\bibfnamefont {B.}~\bibnamefont {Ihrig}}, \bibinfo {author} {\bibfnamefont {L.~N.}\ \bibnamefont {Mihaila}},\ and\ \bibinfo {author} {\bibfnamefont {M.~M.}\ \bibnamefont {Scherer}},\ }\bibfield  {title} {\bibinfo {title} {Critical behavior of dirac fermions from perturbative renormalization},\ }\href {https://doi.org/10.1103/PhysRevB.98.125109} {\bibfield  {journal} {\bibinfo  {journal} {Phys. Rev. B}\ }\textbf {\bibinfo {volume} {98}},\ \bibinfo {pages} {125109} (\bibinfo {year} {2018})}\BibitemShut {NoStop}%
\bibitem [{\citenamefont {Erramilli}\ \emph {et~al.}(2023)\citenamefont {Erramilli}, \citenamefont {Iliesiu}, \citenamefont {Kravchuk}, \citenamefont {Liu}, \citenamefont {Poland},\ and\ \citenamefont {Simmons-Duffin}}]{Erramilli:2022kgp}%
  \BibitemOpen
  \bibfield  {author} {\bibinfo {author} {\bibfnamefont {R.~S.}\ \bibnamefont {Erramilli}}, \bibinfo {author} {\bibfnamefont {L.~V.}\ \bibnamefont {Iliesiu}}, \bibinfo {author} {\bibfnamefont {P.}~\bibnamefont {Kravchuk}}, \bibinfo {author} {\bibfnamefont {A.}~\bibnamefont {Liu}}, \bibinfo {author} {\bibfnamefont {D.}~\bibnamefont {Poland}},\ and\ \bibinfo {author} {\bibfnamefont {D.}~\bibnamefont {Simmons-Duffin}},\ }\bibfield  {title} {\bibinfo {title} {{The Gross-Neveu-Yukawa archipelago}},\ }\href {https://doi.org/10.1007/JHEP02(2023)036} {\bibfield  {journal} {\bibinfo  {journal} {JHEP}\ }\textbf {\bibinfo {volume} {02}},\ \bibinfo {pages} {036}},\ \Eprint {https://arxiv.org/abs/2210.02492} {arXiv:2210.02492 [hep-th]} \BibitemShut {NoStop}%
\bibitem [{\citenamefont {Wang}\ and\ \citenamefont {Meng}(2023)}]{PhysRevB.108.L121112}%
  \BibitemOpen
  \bibfield  {author} {\bibinfo {author} {\bibfnamefont {T.-T.}\ \bibnamefont {Wang}}\ and\ \bibinfo {author} {\bibfnamefont {Z.~Y.}\ \bibnamefont {Meng}},\ }\bibfield  {title} {\bibinfo {title} {Quantum monte carlo calculation of critical exponents of the gross-neveu-yukawa on a two-dimensional fermion lattice model},\ }\href {https://doi.org/10.1103/PhysRevB.108.L121112} {\bibfield  {journal} {\bibinfo  {journal} {Phys. Rev. B}\ }\textbf {\bibinfo {volume} {108}},\ \bibinfo {pages} {L121112} (\bibinfo {year} {2023})}\BibitemShut {NoStop}%
\bibitem [{sup()}]{supplementary}%
  \BibitemOpen
  \href@noop {} {}\bibinfo {note} {See Supplementary Materials for details.}\BibitemShut {Stop}%
\bibitem [{\citenamefont {Wetterich}(1993)}]{Wetterich:1992yh}%
  \BibitemOpen
  \bibfield  {author} {\bibinfo {author} {\bibfnamefont {C.}~\bibnamefont {Wetterich}},\ }\bibfield  {title} {\bibinfo {title} {{Exact evolution equation for the effective potential}},\ }\href {https://doi.org/10.1016/0370-2693(93)90726-X} {\bibfield  {journal} {\bibinfo  {journal} {Phys. Lett. B}\ }\textbf {\bibinfo {volume} {301}},\ \bibinfo {pages} {90} (\bibinfo {year} {1993})},\ \Eprint {https://arxiv.org/abs/1710.05815} {arXiv:1710.05815 [hep-th]} \BibitemShut {NoStop}%
\bibitem [{\citenamefont {Berges}\ \emph {et~al.}(2002)\citenamefont {Berges}, \citenamefont {Tetradis},\ and\ \citenamefont {Wetterich}}]{Berges:2000ew}%
  \BibitemOpen
  \bibfield  {author} {\bibinfo {author} {\bibfnamefont {J.}~\bibnamefont {Berges}}, \bibinfo {author} {\bibfnamefont {N.}~\bibnamefont {Tetradis}},\ and\ \bibinfo {author} {\bibfnamefont {C.}~\bibnamefont {Wetterich}},\ }\bibfield  {title} {\bibinfo {title} {{Nonperturbative renormalization flow in quantum field theory and statistical physics}},\ }\href {https://doi.org/10.1016/S0370-1573(01)00098-9} {\bibfield  {journal} {\bibinfo  {journal} {Phys. Rept.}\ }\textbf {\bibinfo {volume} {363}},\ \bibinfo {pages} {223} (\bibinfo {year} {2002})},\ \Eprint {https://arxiv.org/abs/hep-ph/0005122} {arXiv:hep-ph/0005122} \BibitemShut {NoStop}%
\bibitem [{\citenamefont {Pawlowski}(2007)}]{Pawlowski:2005xe}%
  \BibitemOpen
  \bibfield  {author} {\bibinfo {author} {\bibfnamefont {J.~M.}\ \bibnamefont {Pawlowski}},\ }\bibfield  {title} {\bibinfo {title} {{Aspects of the functional renormalisation group}},\ }\href {https://doi.org/10.1016/j.aop.2007.01.007} {\bibfield  {journal} {\bibinfo  {journal} {Annals Phys.}\ }\textbf {\bibinfo {volume} {322}},\ \bibinfo {pages} {2831} (\bibinfo {year} {2007})},\ \Eprint {https://arxiv.org/abs/hep-th/0512261} {arXiv:hep-th/0512261} \BibitemShut {NoStop}%
\bibitem [{\citenamefont {Baldazzi}\ \emph {et~al.}(2021)\citenamefont {Baldazzi}, \citenamefont {Percacci},\ and\ \citenamefont {Zambelli}}]{Zambelli:2021}%
  \BibitemOpen
  \bibfield  {author} {\bibinfo {author} {\bibfnamefont {A.}~\bibnamefont {Baldazzi}}, \bibinfo {author} {\bibfnamefont {R.}~\bibnamefont {Percacci}},\ and\ \bibinfo {author} {\bibfnamefont {L.}~\bibnamefont {Zambelli}},\ }\bibfield  {title} {\bibinfo {title} {Functional renormalization and the $\overline{\mathrm{ms}}$ scheme},\ }\href {https://doi.org/10.1103/PhysRevD.103.076012} {\bibfield  {journal} {\bibinfo  {journal} {Phys. Rev. D}\ }\textbf {\bibinfo {volume} {103}},\ \bibinfo {pages} {076012} (\bibinfo {year} {2021})}\BibitemShut {NoStop}%
\bibitem [{\citenamefont {Hofling}\ \emph {et~al.}(2002)\citenamefont {Hofling}, \citenamefont {Nowak},\ and\ \citenamefont {Wetterich}}]{Hofling:2002hj}%
  \BibitemOpen
  \bibfield  {author} {\bibinfo {author} {\bibfnamefont {F.}~\bibnamefont {Hofling}}, \bibinfo {author} {\bibfnamefont {C.}~\bibnamefont {Nowak}},\ and\ \bibinfo {author} {\bibfnamefont {C.}~\bibnamefont {Wetterich}},\ }\bibfield  {title} {\bibinfo {title} {{Phase transition and critical behavior of the D = 3 Gross-Neveu model}},\ }\href {https://doi.org/10.1103/PhysRevB.66.205111} {\bibfield  {journal} {\bibinfo  {journal} {Phys. Rev. B}\ }\textbf {\bibinfo {volume} {66}},\ \bibinfo {pages} {205111} (\bibinfo {year} {2002})},\ \Eprint {https://arxiv.org/abs/cond-mat/0203588} {arXiv:cond-mat/0203588} \BibitemShut {NoStop}%
\bibitem [{\citenamefont {Scherer}\ \emph {et~al.}(2013)\citenamefont {Scherer}, \citenamefont {Braun},\ and\ \citenamefont {Gies}}]{Scherer:2012fjq}%
  \BibitemOpen
  \bibfield  {author} {\bibinfo {author} {\bibfnamefont {D.~D.}\ \bibnamefont {Scherer}}, \bibinfo {author} {\bibfnamefont {J.}~\bibnamefont {Braun}},\ and\ \bibinfo {author} {\bibfnamefont {H.}~\bibnamefont {Gies}},\ }\bibfield  {title} {\bibinfo {title} {{Many-flavor Phase Diagram of the (2+1)d Gross-Neveu Model at Finite Temperature}},\ }\href {https://doi.org/10.1088/1751-8113/46/28/285002} {\bibfield  {journal} {\bibinfo  {journal} {J. Phys. A}\ }\textbf {\bibinfo {volume} {46}},\ \bibinfo {pages} {285002} (\bibinfo {year} {2013})},\ \Eprint {https://arxiv.org/abs/1212.4624} {arXiv:1212.4624 [hep-ph]} \BibitemShut {NoStop}%
\bibitem [{\citenamefont {Janssen}\ and\ \citenamefont {Herbut}(2014{\natexlab{b}})}]{janssen2014antiferromagnetic}%
  \BibitemOpen
  \bibfield  {author} {\bibinfo {author} {\bibfnamefont {L.}~\bibnamefont {Janssen}}\ and\ \bibinfo {author} {\bibfnamefont {I.~F.}\ \bibnamefont {Herbut}},\ }\bibfield  {title} {\bibinfo {title} {Antiferromagnetic critical point on graphene's honeycomb lattice: A functional renormalization group approach},\ }\href@noop {} {\bibfield  {journal} {\bibinfo  {journal} {Physical Review B}\ }\textbf {\bibinfo {volume} {89}},\ \bibinfo {pages} {205403} (\bibinfo {year} {2014}{\natexlab{b}})}\BibitemShut {NoStop}%
\bibitem [{\citenamefont {Classen}\ \emph {et~al.}(2017)\citenamefont {Classen}, \citenamefont {Herbut},\ and\ \citenamefont {Scherer}}]{classen2017fluctuation}%
  \BibitemOpen
  \bibfield  {author} {\bibinfo {author} {\bibfnamefont {L.}~\bibnamefont {Classen}}, \bibinfo {author} {\bibfnamefont {I.~F.}\ \bibnamefont {Herbut}},\ and\ \bibinfo {author} {\bibfnamefont {M.~M.}\ \bibnamefont {Scherer}},\ }\bibfield  {title} {\bibinfo {title} {Fluctuation-induced continuous transition and quantum criticality in dirac semimetals},\ }\href@noop {} {\bibfield  {journal} {\bibinfo  {journal} {Physical Review B}\ }\textbf {\bibinfo {volume} {96}},\ \bibinfo {pages} {115132} (\bibinfo {year} {2017})}\BibitemShut {NoStop}%
\bibitem [{\citenamefont {Tolosa-Sime{\'o}n}\ \emph {et~al.}(2025)\citenamefont {Tolosa-Sime{\'o}n}, \citenamefont {Classen},\ and\ \citenamefont {Scherer}}]{Tolosa-Simeon:2025fot}%
  \BibitemOpen
  \bibfield  {author} {\bibinfo {author} {\bibfnamefont {M.}~\bibnamefont {Tolosa-Sime{\'o}n}}, \bibinfo {author} {\bibfnamefont {L.}~\bibnamefont {Classen}},\ and\ \bibinfo {author} {\bibfnamefont {M.~M.}\ \bibnamefont {Scherer}},\ }\bibfield  {title} {\bibinfo {title} {{Relativistic Mott transitions, quantum criticality, and finite-temperature effects in tunable Dirac materials from functional renormalization}},\ }\href@noop {} {\  (\bibinfo {year} {2025})},\ \Eprint {https://arxiv.org/abs/2503.04911} {arXiv:2503.04911 [cond-mat.str-el]} \BibitemShut {NoStop}%
\bibitem [{\citenamefont {Litim}(2001)}]{Litim:2001up}%
  \BibitemOpen
  \bibfield  {author} {\bibinfo {author} {\bibfnamefont {D.~F.}\ \bibnamefont {Litim}},\ }\bibfield  {title} {\bibinfo {title} {{Optimized renormalization group flows}},\ }\href {https://doi.org/10.1103/PhysRevD.64.105007} {\bibfield  {journal} {\bibinfo  {journal} {Phys. Rev. D}\ }\textbf {\bibinfo {volume} {64}},\ \bibinfo {pages} {105007} (\bibinfo {year} {2001})},\ \Eprint {https://arxiv.org/abs/hep-th/0103195} {arXiv:hep-th/0103195} \BibitemShut {NoStop}%
\bibitem [{\citenamefont {D'Attanasio}\ and\ \citenamefont {Morris}(1997)}]{DAttanasio:1997yph}%
  \BibitemOpen
  \bibfield  {author} {\bibinfo {author} {\bibfnamefont {M.}~\bibnamefont {D'Attanasio}}\ and\ \bibinfo {author} {\bibfnamefont {T.~R.}\ \bibnamefont {Morris}},\ }\bibfield  {title} {\bibinfo {title} {{Large N and the renormalization group}},\ }\href {https://doi.org/10.1016/S0370-2693(97)00866-6} {\bibfield  {journal} {\bibinfo  {journal} {Phys. Lett. B}\ }\textbf {\bibinfo {volume} {409}},\ \bibinfo {pages} {363} (\bibinfo {year} {1997})},\ \Eprint {https://arxiv.org/abs/hep-th/9704094} {arXiv:hep-th/9704094} \BibitemShut {NoStop}%
\bibitem [{\citenamefont {Herbut}\ \emph {et~al.}(2009)\citenamefont {Herbut}, \citenamefont {Juri{\v{c}}i{\'c}},\ and\ \citenamefont {Roy}}]{herbut2009theory}%
  \BibitemOpen
  \bibfield  {author} {\bibinfo {author} {\bibfnamefont {I.~F.}\ \bibnamefont {Herbut}}, \bibinfo {author} {\bibfnamefont {V.}~\bibnamefont {Juri{\v{c}}i{\'c}}},\ and\ \bibinfo {author} {\bibfnamefont {B.}~\bibnamefont {Roy}},\ }\bibfield  {title} {\bibinfo {title} {Theory of interacting electrons on the honeycomb lattice},\ }\href@noop {} {\bibfield  {journal} {\bibinfo  {journal} {Physical Review B—Condensed Matter and Materials Physics}\ }\textbf {\bibinfo {volume} {79}},\ \bibinfo {pages} {085116} (\bibinfo {year} {2009})}\BibitemShut {NoStop}%
\bibitem [{\citenamefont {Zamolodchikov}(1986)}]{Zamolodchikov:1986gt}%
  \BibitemOpen
  \bibfield  {author} {\bibinfo {author} {\bibfnamefont {A.~B.}\ \bibnamefont {Zamolodchikov}},\ }\bibfield  {title} {\bibinfo {title} {{Irreversibility of the Flux of the Renormalization Group in a 2D Field Theory}},\ }\href@noop {} {\bibfield  {journal} {\bibinfo  {journal} {JETP Lett.}\ }\textbf {\bibinfo {volume} {43}},\ \bibinfo {pages} {730} (\bibinfo {year} {1986})}\BibitemShut {NoStop}%
\bibitem [{\citenamefont {Cardy}(1988)}]{Cardy:1988cwa}%
  \BibitemOpen
  \bibfield  {author} {\bibinfo {author} {\bibfnamefont {J.~L.}\ \bibnamefont {Cardy}},\ }\bibfield  {title} {\bibinfo {title} {{Is There a c Theorem in Four-Dimensions?}},\ }\href {https://doi.org/10.1016/0370-2693(88)90054-8} {\bibfield  {journal} {\bibinfo  {journal} {Phys. Lett. B}\ }\textbf {\bibinfo {volume} {215}},\ \bibinfo {pages} {749} (\bibinfo {year} {1988})}\BibitemShut {NoStop}%
\bibitem [{\citenamefont {Komargodski}\ and\ \citenamefont {Schwimmer}(2011)}]{Komargodski:2011vj}%
  \BibitemOpen
  \bibfield  {author} {\bibinfo {author} {\bibfnamefont {Z.}~\bibnamefont {Komargodski}}\ and\ \bibinfo {author} {\bibfnamefont {A.}~\bibnamefont {Schwimmer}},\ }\bibfield  {title} {\bibinfo {title} {{On Renormalization Group Flows in Four Dimensions}},\ }\href {https://doi.org/10.1007/JHEP12(2011)099} {\bibfield  {journal} {\bibinfo  {journal} {JHEP}\ }\textbf {\bibinfo {volume} {12}},\ \bibinfo {pages} {099}},\ \Eprint {https://arxiv.org/abs/1107.3987} {arXiv:1107.3987 [hep-th]} \BibitemShut {NoStop}%
\bibitem [{\citenamefont {Klebanov}\ \emph {et~al.}(2011)\citenamefont {Klebanov}, \citenamefont {Pufu},\ and\ \citenamefont {Safdi}}]{Klebanov:2011gs}%
  \BibitemOpen
  \bibfield  {author} {\bibinfo {author} {\bibfnamefont {I.~R.}\ \bibnamefont {Klebanov}}, \bibinfo {author} {\bibfnamefont {S.~S.}\ \bibnamefont {Pufu}},\ and\ \bibinfo {author} {\bibfnamefont {B.~R.}\ \bibnamefont {Safdi}},\ }\bibfield  {title} {\bibinfo {title} {{F-Theorem without Supersymmetry}},\ }\href {https://doi.org/10.1007/JHEP10(2011)038} {\bibfield  {journal} {\bibinfo  {journal} {JHEP}\ }\textbf {\bibinfo {volume} {10}},\ \bibinfo {pages} {038}},\ \Eprint {https://arxiv.org/abs/1105.4598} {arXiv:1105.4598 [hep-th]} \BibitemShut {NoStop}%
\bibitem [{\citenamefont {Casini}\ and\ \citenamefont {Huerta}(2012)}]{Casini:2012ei}%
  \BibitemOpen
  \bibfield  {author} {\bibinfo {author} {\bibfnamefont {H.}~\bibnamefont {Casini}}\ and\ \bibinfo {author} {\bibfnamefont {M.}~\bibnamefont {Huerta}},\ }\bibfield  {title} {\bibinfo {title} {{On the RG running of the entanglement entropy of a circle}},\ }\href {https://doi.org/10.1103/PhysRevD.85.125016} {\bibfield  {journal} {\bibinfo  {journal} {Phys. Rev. D}\ }\textbf {\bibinfo {volume} {85}},\ \bibinfo {pages} {125016} (\bibinfo {year} {2012})},\ \Eprint {https://arxiv.org/abs/1202.5650} {arXiv:1202.5650 [hep-th]} \BibitemShut {NoStop}%
\bibitem [{\citenamefont {Casini}\ \emph {et~al.}(2017)\citenamefont {Casini}, \citenamefont {Test{\'e}},\ and\ \citenamefont {Torroba}}]{Casini:2017vbe}%
  \BibitemOpen
  \bibfield  {author} {\bibinfo {author} {\bibfnamefont {H.}~\bibnamefont {Casini}}, \bibinfo {author} {\bibfnamefont {E.}~\bibnamefont {Test{\'e}}},\ and\ \bibinfo {author} {\bibfnamefont {G.}~\bibnamefont {Torroba}},\ }\bibfield  {title} {\bibinfo {title} {{Markov Property of the Conformal Field Theory Vacuum and the a Theorem}},\ }\href {https://doi.org/10.1103/PhysRevLett.118.261602} {\bibfield  {journal} {\bibinfo  {journal} {Phys. Rev. Lett.}\ }\textbf {\bibinfo {volume} {118}},\ \bibinfo {pages} {261602} (\bibinfo {year} {2017})},\ \Eprint {https://arxiv.org/abs/1704.01870} {arXiv:1704.01870 [hep-th]} \BibitemShut {NoStop}%
\bibitem [{Note2()}]{Note2}%
  \BibitemOpen
  \bibinfo {note} {The contribution of the divergent zero mode is excluded using dimensional regularization.}\BibitemShut {Stop}%
\bibitem [{\citenamefont {Braun}(2009)}]{Braun:2009ewx}%
  \BibitemOpen
  \bibfield  {author} {\bibinfo {author} {\bibfnamefont {J.}~\bibnamefont {Braun}},\ }\bibfield  {title} {\bibinfo {title} {{The QCD Phase Boundary from Quark-Gluon Dynamics}},\ }\href {https://doi.org/10.1140/epjc/s10052-009-1136-6} {\bibfield  {journal} {\bibinfo  {journal} {Eur. Phys. J. C}\ }\textbf {\bibinfo {volume} {64}},\ \bibinfo {pages} {459} (\bibinfo {year} {2009})},\ \Eprint {https://arxiv.org/abs/0810.1727} {arXiv:0810.1727 [hep-ph]} \BibitemShut {NoStop}%
\bibitem [{\citenamefont {Classen}\ \emph {et~al.}(2016)\citenamefont {Classen}, \citenamefont {Herbut}, \citenamefont {Janssen},\ and\ \citenamefont {Scherer}}]{Classen:2015mar}%
  \BibitemOpen
  \bibfield  {author} {\bibinfo {author} {\bibfnamefont {L.}~\bibnamefont {Classen}}, \bibinfo {author} {\bibfnamefont {I.~F.}\ \bibnamefont {Herbut}}, \bibinfo {author} {\bibfnamefont {L.}~\bibnamefont {Janssen}},\ and\ \bibinfo {author} {\bibfnamefont {M.~M.}\ \bibnamefont {Scherer}},\ }\bibfield  {title} {\bibinfo {title} {{Competition of density waves and quantum multicritical behavior in Dirac materials from functional renormalization}},\ }\href {https://doi.org/10.1103/PhysRevB.93.125119} {\bibfield  {journal} {\bibinfo  {journal} {Phys. Rev. B}\ }\textbf {\bibinfo {volume} {93}},\ \bibinfo {pages} {125119} (\bibinfo {year} {2016})},\ \Eprint {https://arxiv.org/abs/1510.09003} {arXiv:1510.09003 [cond-mat.str-el]} \BibitemShut {NoStop}%
\bibitem [{\citenamefont {De~Polsi}\ \emph {et~al.}(2020)\citenamefont {De~Polsi}, \citenamefont {Balog}, \citenamefont {Tissier},\ and\ \citenamefont {Wschebor}}]{DePolsi:2020pjk}%
  \BibitemOpen
  \bibfield  {author} {\bibinfo {author} {\bibfnamefont {G.}~\bibnamefont {De~Polsi}}, \bibinfo {author} {\bibfnamefont {I.}~\bibnamefont {Balog}}, \bibinfo {author} {\bibfnamefont {M.}~\bibnamefont {Tissier}},\ and\ \bibinfo {author} {\bibfnamefont {N.}~\bibnamefont {Wschebor}},\ }\bibfield  {title} {\bibinfo {title} {{Precision calculation of critical exponents in the $O(N)$ universality classes with the nonperturbative renormalization group}},\ }\href {https://doi.org/10.1103/PhysRevE.101.042113} {\bibfield  {journal} {\bibinfo  {journal} {Phys. Rev. E}\ }\textbf {\bibinfo {volume} {101}},\ \bibinfo {pages} {042113} (\bibinfo {year} {2020})},\ \Eprint {https://arxiv.org/abs/2001.07525} {arXiv:2001.07525 [cond-mat.stat-mech]} \BibitemShut {NoStop}%
\end{thebibliography}%
\clearpage
\onecolumngrid

\appendix   

\section*{Supplementary materials: epsilon expansion}

\section{Beta functions}
\label{appx:RG}

To renormalize the model, we supplement (\ref{action}) with the following counterterms
\begin{align}
 S_\text{c.t.}&=\int_\mathcal{B} ~ \Big({\delta Z_\phi\over 2}(\partial\phi)^2 
 + \frac{\delta Z_\chi}{2}(\partial\chi)^2
 + \delta Z_\psi\bar{\psi}\slashed{\partial}\psi  
 \nonumber \\
 &+ \delta Z_{\lambda_\phi} {\lambda_\phi \mu^{\epsilon}\over 8}(\phi^2)^2 
 + \delta Z_{\lambda_{\phi\chi}} {\lambda_{\phi\chi}\mu^{\epsilon}\over 4} \, \phi^2\chi^2
 +\frac{\delta Z_{\lambda_\chi}}{8} \, \lambda_\chi\, \mu^\epsilon  \chi^4
 +\delta Z_h \, h\, \mu^\frac{\epsilon}{2} \, \bar{\psi}\psi\chi\Big)  ~,
\end{align}
where $\delta Z_i$ for $i=\phi, \chi, \psi, \lambda_\phi,\lambda_{\phi\chi}, \lambda_\chi, h$ contain an ascending series of poles in $1/\epsilon$; that is, we employ the minimal subtraction scheme. Define $Z_i=1+\delta Z_i$, then the relations between the bare and renormalized parameters are given by
\begin{gather}
		\qquad\phi_0=\sqrt{Z_\phi} \, \phi ~,\qquad\chi_0=\sqrt{Z_\chi} \, \chi  ~, \qquad\psi_0=\sqrt{Z_\psi} \, \psi ~,
        \nonumber\\
		\lambda_{\phi}^0= {Z_{\lambda_\phi}\over Z_\phi^2} \lambda_\phi \,\mu^{\epsilon},\qquad 
        \lambda_{\phi\chi}^0= {Z_{\lambda_{\phi\chi}}\over Z_\phi Z_\chi}\lambda_{\phi\chi}\, \mu^{\epsilon},\qquad \lambda_{\chi}^0= {Z_\lambda\over Z^2_\chi}   \lambda_\chi \, \mu^\epsilon,\qquad
        h_0=  {Z_h\over \sqrt{Z_\chi} Z_\psi } \, h\, \mu^\frac{\epsilon}{2}  ~.
		\label{CTdefs}
\end{gather}
In this appendix, we carry out explicit calculations of the 1-loop Feynman graphs contributing to the renormalization constants $Z_i$. The position space representations of the two-point functions for a massless scalar field and for a Dirac field are given by
\begin{align}
  \langle \phi(x_1) \phi(x_2) \rangle&=
  \langle \chi(x_1) \chi(x_2) \rangle =
  {C_\chi \over |x_{12}|^{d-2}} ~, \quad
  C_\chi={ \Gamma\big( {d-2\over 2}\big) \over 4\pi^{d\over 2} } ~, \quad
  x_{12}=x_1-x_2 ~,
  \nonumber \\
 \langle \psi(x_1) \bar\psi(x_2) \rangle &= 
 C_\psi {\slashed{x}_{12} \over |x_{12}|^{d}} ~, \quad 
 C_\psi={\Gamma\big({d\over 2}\big)\over 2\pi^{d\over 2}}~.
 \label{props}
\end{align}

\begin{figure}[b!]
		\centering
		\includegraphics[width=0.5\columnwidth]{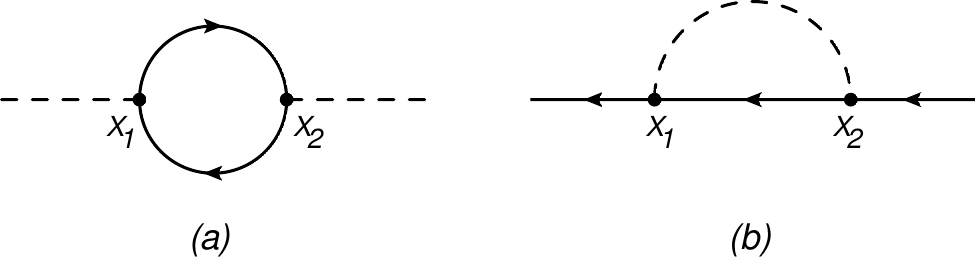}
		\caption{One-loop diagrams contributing to the wave function renormalization for \textbf{(a)} the scalar field $\chi$ and \textbf{(b)} the Dirac field $\psi$. Internal solid lines with arrows correspond to the propagator of the Dirac field, while internal dashed lines represent the propagator of the scalar field $\chi$. External dashed and solid lines denote scalar and Dirac background fields, respectively.}
		\label{fig:wave}
	\end{figure}

The counterterm $\delta Z_\phi$ is trivial at one-loop, while the wave function renormalization of $\chi$ and $\psi$ are associated with the diagrams in Fig.\ref{fig:wave},
\begin{align}
 \text{Fig.}\ref{fig:wave}\text{(a)}&= - {N_f h^2 \mu^{\epsilon} \, C_\psi^2 \over 2} \int_\mathcal{B} \int_\mathcal{B} \chi(x_1) \chi(x_2)\text{tr} \Big( {\slashed{x}_{12}  \slashed{x}_{21}\over |x_{12}|^{d} |x_{21}|^d}\Big) = - {N_2h^2 \over 32\pi^2 \epsilon} \int_\mathcal{B} \partial\chi \partial\chi + \ldots ~,
 \nonumber
  \\
 \text{Fig.}\ref{fig:wave}\text{(b)}&= h^2 \mu^\epsilon  C_\psi C_\chi   \int_\mathcal{B}  \int_\mathcal{B} {\bar\psi(x_1) \slashed{x}_{12} \psi(x_2) \over (x_{12}^2)^{d-1}}
 =-{h^2 \over 16\pi^2 \epsilon} \int_\mathcal{B}\bar{\psi}\slashed{\partial}\psi + \ldots ~.
 \nonumber
\end{align}
where the trace is done over the spinor indices. Hence,
\begin{align} 
	&Z_\chi=1- \frac{N_2h^2}{16\pi^2\epsilon}+ \ldots  ~, 	\nonumber \\
	& Z_\psi= 1- {h^2\over 16\pi^2 \epsilon} + \ldots ~. \label{wave-renorm}
\end{align}

\begin{figure}[t!]
		\centering
		\includegraphics[width=0.3\columnwidth]{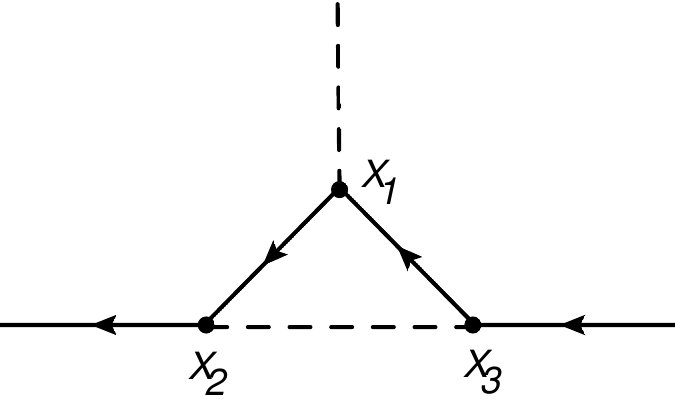}
		\caption{One-loop diagram contributing to the renormalization of the Yukawa coupling $h$. Internal solid lines with arrows correspond to the propagator of the Dirac field $\psi$, while the internal dashed line represents the propagator of the scalar field $\chi$. In contrast, external dashed and solid lines denote scalar and Dirac background fields, respectively.}
		\label{fig:Zeta}
	\end{figure}

Next, we fix $Z_h$ by evaluating the one-loop diagram in Fig. \ref{fig:Zeta},
\begin{equation}
  \text{Fig.}\ref{fig:Zeta}= - h^3 \mu^{3\epsilon\over 2} C_\psi^2 C_\chi \int_\mathcal{B} \int_\mathcal{B}  \int_\mathcal{B} \chi(x_1)
  {\bar\psi(x_2)\slashed{x}_{21} \slashed{x}_{13} \psi(x_3)\over |x_{21}|^d|x_{31}|^d|x_{23}|^{d-2}} ~.
\end{equation} 
The pole in $\epsilon$ of this integral is obtained by expanding the Dirac fields around $x_1$ and subsequently integrating over $x_2$ and $x_3$ using the scalar-fermion propagator merging relation of the form
\begin{align} 
 \int d^d x_3 {\slashed{x}_{31} \over |x_{31}|^{2\Delta_1+1} |x_{23}|^{2\Delta_2}} &=
 { \pi^{d\over 2}  \Gamma\big( {d\over 2} -  \Delta_2\big)  
 \Gamma\big({d\over 2}-\Delta_1+{1\over 2}\big) 
 \Gamma\big(\Delta_1+\Delta_2-{d\over 2}+{1\over 2}\big) \over \Gamma(\Delta_2) \Gamma\big(\Delta_1+{1\over 2}\big) \Gamma\big(d-\Delta_1-\Delta_2+{1\over 2}\big)} 
 \nonumber \\
 &\times{\slashed{x}_{21} \over |x_{12}|^{2(\Delta_1+\Delta_2)-d+1}} ~. \label{mergingFS}
\end{align}
The final expression is given by
\begin{equation}
  \text{Fig.}\ref{fig:Zeta}=   {h^3 \mu^{\epsilon\over 2}\over 8\pi^2\epsilon} \int_\mathcal{B} \bar\psi\psi\chi + \ldots ~.
\end{equation} 
Thus,
\begin{equation}
 Z_h=1 + {h^2 \over 8\pi^2\epsilon} + \ldots ~.
 \label{eta-renorm}
\end{equation}
Substituting (\ref{wave-renorm}) and (\ref{eta-renorm}) into the relation between $h_0$ and $h$ in (\ref{CTdefs}), gives \cite{Zinn-Justin:1991ksq,Moshe:2003xn}
\begin{equation}
 h_0=  {Z_h\over \sqrt{Z_\chi} Z_\psi } \, \mu^\frac{\epsilon}{2} \, h \quad \Rightarrow \quad \beta_{h^2} = -\epsilon h^2+\frac{N_2+6}{16\pi^2}h^4 ~.
\end{equation}
\begin{figure}[h!]
		\centering
		\includegraphics[width=0.5\columnwidth]{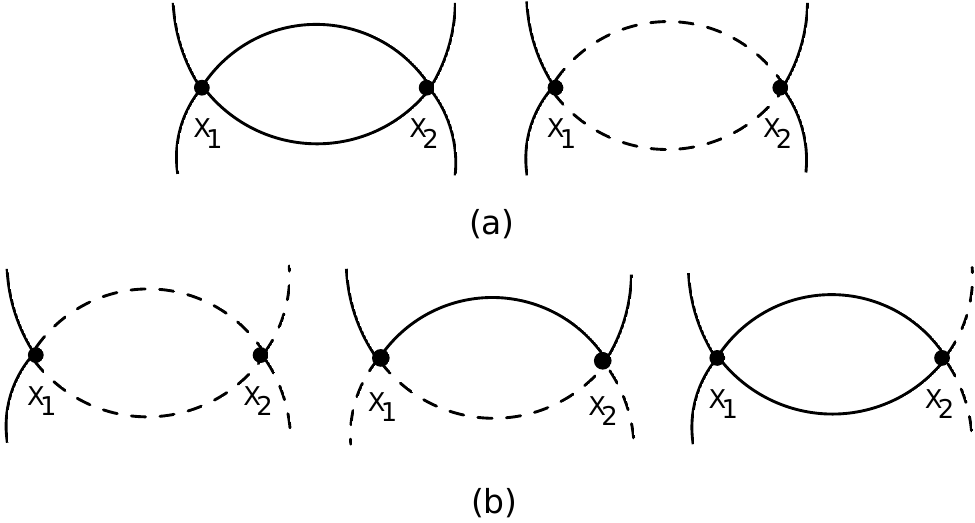}
		\caption{One-loop diagrams contributing to $Z_{\lambda_\phi},Z_{\lambda_{\phi\chi}}$. Internal solid lines correspond to the propagator of $\phi$, while the internal dashed lines represent the propagator of the scalar field $\chi$. External solid and dashed lines denote scalar background fields.}
		\label{fig:Zeta12}
	\end{figure}
To determine $Z_{\lambda_\phi}, Z_{\lambda_{\phi\chi}}$, one has to evaluate the one-loop diagrams shown in Fig. \ref{fig:Zeta12}
\begin{align}
 \text{Fig.}\ref{fig:Zeta12}\text{(a)}&=   {\lambda_\phi^2\mu^{2\epsilon} C_\chi^2\over 128}
 \Big((8N_1+32)\delta_{ac}\delta_{be} + 32 \delta_{ab}\delta_{ce}\Big)
 \int_\mathcal{B} \int_\mathcal{B}   
 {\phi_a(x_1) \phi_c(x_1)\phi_b(x_2) \phi_e(x_2)\over |x_{12}|^{2(d-2)} }
 \nonumber \\
 &+{\lambda_{\phi\chi}^2\mu^{2\epsilon} C_\chi^2\over 16}
 \int_\mathcal{B} \int_\mathcal{B}   
 {\phi^2(x_1) \phi^2(x_2)\over |x_{12}|^{2(d-2)} } ~,
 \nonumber \\
 \text{Fig.}\ref{fig:Zeta12}\text{(b)}&= {3\lambda_{\phi\chi} \lambda_\chi \mu^{2\epsilon} C_\chi^2 \over  8}
 \int_\mathcal{B} \int_\mathcal{B}   
 {\phi^2(x_1) \chi^2(x_2)\over |x_{12}|^{2(d-2)} }
 +{\lambda_{\phi\chi}^2\mu^{2\epsilon} C_\chi^2\over 2}
 \int_\mathcal{B} \int_\mathcal{B}   
 {\phi_a(x_1) \chi(x_1) \phi_a(x_2)\chi(x_2)\over |x_{12}|^{2(d-2)} }
 \nonumber\\
 &+ { (N_1+2) \lambda_\phi \lambda_{\phi\chi}\mu^{2\epsilon} C_\chi^2\over 8}
 \int_\mathcal{B} \int_\mathcal{B}   
 {\chi^2(x_1) \phi^2(x_2)\over |x_{12}|^{2(d-2)}}   ~.
\end{align}
Expanding the background fields around $x_1$ gives
\begin{align}
 \text{Fig.}\ref{fig:Zeta12}\text{(a)}&= {\mu^\epsilon\over 128\pi^2 \epsilon} \big( (N_1+8)\lambda_\phi^2+\lambda_{\phi\chi}^2\big)
 \int_\mathcal{B} (\phi^2)^2+ \ldots ~,
 \nonumber \\ 
 \text{Fig.}\ref{fig:Zeta12}\text{(b)}&=
 {\mu^\epsilon\over 64\pi^2 \epsilon}
 \big( 3 \lambda_{\phi\chi}\lambda_\chi + 4 \lambda_{\phi\chi}^2 + (N_1+2)\lambda_{\phi} \lambda_{\phi\chi}\big)
 \int_\mathcal{B}\chi^2 \phi^2 + \ldots ~.
 \nonumber
\end{align}
Hence,
\begin{align}
    Z_{\lambda_\phi}&=
    1+{\lambda_\phi^{-1}\over 16\pi^2 \epsilon} 
    \big((N_1+8)\lambda_\phi^2+\lambda_{\phi\chi}^2\big)+ \ldots ~,
    \nonumber \\
    Z_{\lambda_{\phi\chi}}&=
    1+ {1\over 16\pi^2 \epsilon}
 \big( 3 \lambda_\chi + 4 \lambda_{\phi\chi} + (N_1+2)\lambda_\phi\big) + \ldots ~.
 \label{g-renorm}
\end{align}
Substituting (\ref{wave-renorm}) and (\ref{g-renorm}) into the relation between bare and renormalized couplings in (\ref{CTdefs}), gives
\begin{align}
    \lambda_{\phi}^0&= {Z_{\lambda_\phi}\over Z_\phi^2} \lambda_\phi \mu^{\epsilon}
    \quad \Rightarrow \quad 
    \beta_{\lambda_\phi} = - \epsilon \lambda_{\phi} + {N_1+8\over 16 \pi^2} \lambda_{\phi}^2
    + {\lambda_{\phi\chi}^2\over 16 \pi^2}  + \ldots ~,\\
    \lambda_{\phi\chi}^0&= {Z_{\lambda_{\phi\chi}}\over Z_\phi Z_\chi}\lambda_{\phi\chi} \, \mu^{\epsilon}
    \quad \Rightarrow \quad 
    \beta_{\lambda_{\phi\chi}} = - \epsilon \lambda_{\phi\chi} + {N_2\over 16\pi^2} \lambda_{\phi\chi} h^2
    +{3\over 16\pi^2}\lambda_\chi \lambda_{\phi\chi} +{\lambda_{\phi\chi}^2\over 4\pi^2}
    +{N_1+2\over 16\pi^2} \lambda_{\phi}\lambda_{\phi\chi} + \ldots ~.
    \nonumber
\end{align}

\begin{figure}[b!]
		\centering
		\includegraphics[width=0.6\columnwidth]{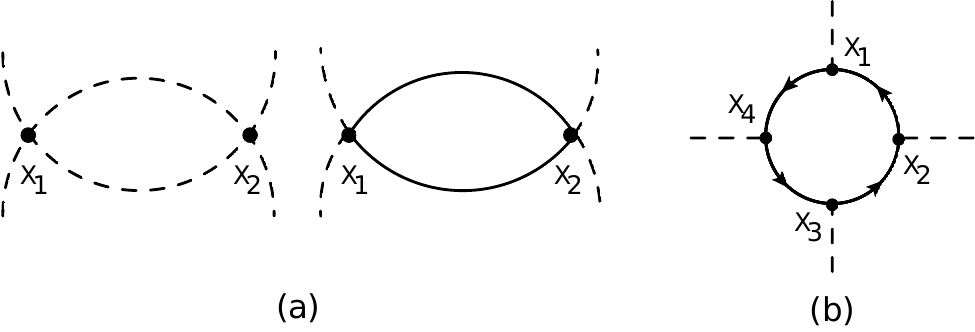}
		\caption{One-loop diagrams contributing to the renormalization of the coupling $\lambda$. Internal solid lines with/without arrows correspond to the propagator of $\psi$/$\phi$, respectively, while the internal dashed lines represent the propagator of the scalar field $\chi$. External dashed lines denote scalar background fields.}
		\label{fig:Zlam}
	\end{figure}

To derive the RG flow for $\lambda_\chi$, one has to consider the one-loop diagrams shown in Fig. \ref{fig:Zlam} 
\begin{align}
 \text{Fig.}\ref{fig:Zlam}\text{(a)}&= {C_\chi^2\over 16}\mu^{2\epsilon} \big(
 9\lambda_\chi^2
 +N_1 \lambda_{\phi\chi}^2\big)\int_\mathcal{B}\int_\mathcal{B} {\chi^2(x_1) \chi^2(x_2)\over |x_{12}|^{2(d-2)}}  ~,
  \\
 \text{Fig.}\ref{fig:Zlam}\text{(b)}&= - {h^4 \mu^{2\epsilon} C_\psi^4 N_f\over 4}  \int_\mathcal{B} \int_\mathcal{B} \int_\mathcal{B} \int_\mathcal{B} \prod_{i=1}^4 \chi(x_i)
  \text{tr} \big( {\slashed{x}_{12} \slashed{x}_{23} \slashed{x}_{34} \slashed{x}_{41} \over |x_{12}|^d|x_{23}|^d|x_{34}|^d|x_{41}|^d} \big)  ~.
   \nonumber
\end{align}
The counterterm $Z_\lambda$ is entirely fixed by the poles in $\epsilon$, therefore we expand the background fields around $x_1$ and get
\begin{align}
 \text{Fig.}\ref{fig:Zlam}\text{(a)}&=  
 {\mu^\epsilon\over 128\pi^2 \epsilon}
 \big(9\lambda_\chi^2 
 + N_1 \lambda_{\phi\chi}^2\big)
 \int_\mathcal{B} \chi^4 +\ldots  ~,
 \nonumber
  \\
 \text{Fig.}\ref{fig:Zlam}\text{(b)}&= -{h^4 \mu^\epsilon N_2\over 32 \pi^2 \epsilon} \int_\mathcal{B} \chi^4  +\ldots  ~.
\end{align}
To get the above expression for \text{Fig.} \ref{fig:Zlam}\text{(b)}, we used (\ref{mergingFS}) and the merging relation for the pair of Dirac propagators,
\begin{align}
 \int d^d x_3 {\slashed{x}_{13}\slashed{x}_{32} \over |x_{32}|^{2\Delta_1+1} |x_{13}|^{2\Delta_2+1}} &=
 {- \pi^{d\over 2}  \Gamma\big( \Delta_1 + \Delta_2-{d\over 2}\big)  \Gamma\big({d\over 2}-\Delta_1+{1\over 2}\big) \Gamma\big({d\over 2}-\Delta_2+{1\over 2}\big) \over \Gamma(d-\Delta_1-\Delta_2) \Gamma\big(\Delta_1+{1\over 2}\big) \Gamma\big(\Delta_2+{1\over 2}\big)} 
 \nonumber \\
 &\times{\mathbb{I} \over |x_{12}|^{2(\Delta_1+\Delta_2)-d}} ~.
 \label{mergingFF}
\end{align}
Hence,
\begin{equation}
 Z_\lambda=1 + {9\lambda_\chi \over 16 \pi^2 \epsilon} 
 + {N_1\over 16\pi^2\epsilon} {\lambda_{\phi\chi}^2\over \lambda_\chi} 
 - {  N_2\over 4 \pi^2  \epsilon} \,
 {h^4\over \lambda_\chi} + \ldots ~.
\end{equation}
As a result, we obtain the following RG flow for $\lambda$,
	\begin{gather}
		\lambda_\chi^0= {Z_\lambda\over Z^2_\chi} ~ \mu^\epsilon  \lambda_\chi \quad \Rightarrow \quad \beta_{\lambda_\chi} = -\epsilon\lambda_\chi +\frac{1}{8\pi^2}\Big(\frac{9}{2}\lambda_\chi^2 +{N_1\over 2}\lambda_{\phi\chi}^2 + N_2\lambda_\chi h^2-2N_2h^4\Big) ~.
	\end{gather}

\section{Thermal Masses}
\label{massT}
It is straightforward to generalize \eqref{props} to the thermal case
\begin{align}
 \langle \chi(\tau,\vec x) \chi(0)\rangle_T &= \sum_{n=-\infty}^\infty
    {C_\chi\over  \big((\tau+n\beta)^2 + \vec x^2\big)^{d-2\over 2}}
    \nonumber \\
 \langle \psi(\tau,\vec x) \bar\psi(0) \rangle_T &=
 {-i\over \beta}\sum_{n=-\infty}^\infty
  \int {d^{d-1}\vec k\over (2\pi)^{d-1}}
 {\slashed{k}\over k^2} e^{i(k^0\tau+\vec k \cdot \vec x)}~, \quad
 k^0={(2n+1)\pi\over \beta} ~.
\end{align}
By construction, the correlators for (fermion)scalar fields satisfy the appropriate Green's equation and obey (anti)periodic boundary conditions along the thermal circle. 

Using the scalar thermal two-point function, one can evaluate the following thermal expectation value in free field theory\footnote{The contribution of the divergent zero mode is excluded using dimensional regularization.}
\begin{equation}
   L_S=\langle \chi^2\rangle_T = 2 \, C_\chi \, \beta^{2-d} \sum_{n=1}^\infty
    {1 \over n^{d-2}}= {\Gamma\big({d-2\over 2}\big)\zeta(d-2)\over 2\pi^{d\over 2}} \, \beta^{2-d}= {T^2\over 12} + \mathcal{O}(d-4)~.
\end{equation}
Similarly,
\begin{equation}
 \int d\tau \, d^{d-1}\vec x \, \langle \bar\psi\psi(\tau,\vec x) \, \bar\psi\psi(0) \rangle_T ={N_2\over \beta}\sum_{n=-\infty}^\infty
  \int {d^{d-1}\vec k\over (2\pi)^{d-1}}{1\over {(2n+1)^2\pi^2\over\beta^2} + \vec k^2}~.
\end{equation}
Introducing spherical coordinates to evaluate the integral over $\vec k$, yields
\begin{equation}
 L_F=\int d\tau \, d^{d-1}\vec x \, \langle \bar\psi\psi(\tau,\vec x) \, \bar\psi\psi(0) \rangle_T =-N_2\beta^{2-d}{2^{2-d}\pi^{d-3\over 2}\over \Gamma\big({d-1\over 2}\big) \sin\big({d\pi\over 2}\big)}\sum_{n=0}^\infty (2n+1)^{d-3}
  ~.
\end{equation}
Hence,
\begin{equation}
 L_F=\int d\tau \, d^{d-1}\vec x \, \langle \bar\psi\psi(\tau,\vec x) \, \bar\psi\psi(0) \rangle_T =N_2\beta^{2-d}{(1-2^{3-d})\pi^{d-3\over 2} \zeta(3-d)\over 2\, \Gamma\big({d-1\over 2}\big) \sin\big({d\pi\over 2}\big)}
 =-{N_2 \over 24} \, T^2 + \mathcal{O}(d-4)~.
\end{equation}
Diagrammatically, $L_\text{S}$ and $L_\text{F}$ can be represented as scalar and fermionic one-loop graphs, respectively. In the weakly coupled QFT we focus on, they generate leading-order quadratic corrections to the effective potential at finite temperature. These corrections correspond to thermally induced mass terms, commonly referred to as thermal masses. Specifically, at one-loop order, the thermally induced masses of $\phi$ and $\chi$ are given by
\begin{align}
 m_\phi^2(T)\Big|_\text{1-loop}&=\Big((N_1+2)\lambda_\phi + \lambda_{\phi\chi} \Big) {L_S\over 2} ~,
 \nonumber \\
 m_\chi^2(T)\Big|_\text{1-loop}&= \Big(3\lambda_\chi + N_1\lambda_{\phi\chi}\Big) {L_S\over 2} - h^2 \, L_F ~.
\end{align}

\section*{Supplementary materials: FRG}

\section{IR regulator, threshold functions and thermal factors}
We define the bosonic and fermionic IR regulators in the following manner
\begin{equation}
    \Delta S_k^B [\phi]= \int \frac{d^d q}{(2\pi)^d} \phi(-q) R^{B}_{k}(q^2) \phi(q), \quad \Delta S_k^\psi [\psi,\bar{\psi}]= \int \frac{d^d q}{(2\pi)^d} \bar{\psi}(q) R^\psi_{k}(q) \psi(q), \quad
    R_k = \left(\begin{smallmatrix}
        R^\phi_k & 0 & 0 & 0\\
        0 & R^\chi_k & 0 & 0\\
        0 & 0 & 0 & - R^{\psi\,T}_k\\
        0 & 0 & R^\psi_k &  0 \\
    \end{smallmatrix}\right),
\end{equation}
reminiscent of the form of the kinetic terms in momentum space $\Delta S_{\text{kin}}^B [\phi]= \int \frac{d^d q}{(2\pi)^d} \frac{Z_{\phi}}{2} \phi(-q)\, q^2 \phi(q)$, $ \Delta S_{\text{kin}}^\psi [\psi,\bar{\psi}]= \int \frac{d^d q}{(2\pi)^d} 
Z_\psi \bar{\psi}(q) i\slashed{q} \psi(q)$.
The renormalized dimensionless IR regulators $r_k^B$ and $r_k^\psi$ are defined as
\begin{equation}
    R_k^{B}(q^2) = Z_{B,k} q^2r_B \left(\frac{q^2}{k^2}\right), \quad R_k^\psi(q) = Z_{\psi,k} i \slashed{q} r_\psi\left(\frac{q^2}{k^2}\right), \quad B\in \{\phi, \chi\}.
\end{equation}
Derivatives with respect to $t = \ln k/\Lambda$ read
\begin{equation}
    \partial_t R^B_k = Z_{B,k} q^2 (-2yr_B'-\eta_B r_B), \quad \partial_t R^\psi_k = Z_{\psi,k} i \slashed{q} (-2y r'_\psi -\eta_\psi r_\psi), 
\end{equation}
where $\eta_i = - \partial_t \ln Z_{i,k}$. We will encounter the $\tilde{\partial}_t$ derivative, which is defined to act on the $k$-scale dependence only of the regulators. Thus, one can rewrite it as 
\begin{equation} \label{eq: dttilde}
    \tilde{\partial}_t = \sum_{\Phi \in \{\phi,\chi,\psi\}} \int_0^\infty dy\, (-2y\,r_{\Phi}'(y)-\eta_{\Phi}r_\Phi (y)) \frac{\delta}{\delta r_{\Phi}(y)}.
\end{equation}
We employ the linear regulator \cite{Litim:2001up}
\begin{equation}
    r_B(y) = \left(\frac{1}{y} - 1\right) \Theta(1-y), \quad r_\psi (y) =  \left(\frac{1}{\sqrt{y}} - 1\right) \Theta(1-y),
\end{equation}
with the property $y(1+r_B) = y (1+r_\psi)^2$. Some preliminary definitions 
\begin{flalign}
    &\omega_{\phi} = u_k^{(1,0)} + 2\bar{\rho}_\phi u_k^{(2,0)},\quad \omega_{\chi} = u_k^{(0,1)} + 2\bar{\rho}_\chi u_k^{(0,2)}, \quad \omega_{\phi \chi} = 2 \sqrt{\bar{\rho}_\phi \bar{\rho}_\chi} u^{(1,1)}, \quad \omega_\psi = 2\bar{\rho}_\chi \bar{h}^2, \quad \tau \equiv 2\pi T/k, \notag\\
    &\omega_{\chi\chi\chi} = \sqrt{2\bar{\rho}_\chi} (3 u^{(0,2)} + 2 \bar{\rho}_{\chi} u^{(0,3)}),\quad \omega_{\chi\chi\phi} = \sqrt{2\bar{\rho}_\phi}(u^{(1,1)} + 2\bar{\rho}_\chi u^{(1,2)}),\quad \omega_{\chi\phi\phi} = \sqrt{2\bar{\rho}_\chi} (u^{(1,1)} + 2\bar{\rho}_\phi u^{(2,1)}).
\end{flalign}
The threshold functions evaluated for the linear regulator read 
\begin{equation}
    I_R (\omega_1,\omega_2,\omega_3) = \frac{4v_d}{d} \frac{1 + \omega_1}{(1+\omega_1)(1+\omega_2)-\omega_3^2},\quad I_G (\omega) = \frac{4v_d}{d} \frac{1}{1+\omega}, \quad I_F (\omega) = \frac{4v_d}{d} \frac{1}{1+\omega}, 
\end{equation}
\begin{equation}\label{eq: mF}
    m_{F}(\omega,\eta,\tau) = \left(\frac{1-\eta }{d-2}\frac{1}{(1+\omega)^3}-\frac{1-\eta }{2 d-4}\frac{1}{(1+\omega)^2}\right) t_F(\tau ) +
   \left(\frac{2}{(1+\omega)^4}-\frac{1}{(1+\omega)^3}-\frac{1/4}{(1+\omega)^2}\right) s_F(\tau ),
\end{equation}
\begin{multline} \label{eq: lchichi}
l_{(nm)}^{\chi\chi} = \frac{2}{d} \frac{\left(h (1 + \omega_\phi)\right)^m}{\left((1 + \omega_\phi)(1 + \omega_\chi)-\omega_{\phi\chi}^2\right)^m (1 + \omega_\psi)^n} \times \\
\times \Bigg[ \frac{m \left(s_B(\tau) - \frac{\eta_\phi}{2 + d} \hat{s}_B(\tau)\right)\omega_{\phi\chi}^2}{(1 + \omega_\phi)\left((1 + \omega_\phi)(1 + \omega_\chi)-\omega_{\phi\chi}^2\right)} + \frac{m \left(s_B(\tau) - \frac{\eta_\chi}{2 + d} \hat{s}_B (\tau)\right)(1 + \omega_\phi)}{(1 + \omega_\phi)(1 + \omega_\chi)-\omega_{\phi\chi}^2} + \frac{n \left(s_F(\tau) - \frac{\eta_\psi}{1 + d} \hat{s}_F(\tau)\right)}{1 + \omega_\psi} \Bigg],
\end{multline}
\begin{multline} \label{eq: lphiphi}
l_{(nm)}^{\phi\phi} = 
\frac{2}{d}\frac{(h \omega_{\phi\chi})^m}{\left((1 + \omega_\phi)(1 + \omega_\chi)-\omega_{\phi\chi}^2\right)^m (1 + \omega_\psi)^n} \times \\
\times \Bigg[ 
\frac{m \left(s_B(\tau) - \frac{\eta_\phi}{2 + d} \hat{s}_B(\tau)\right)(1 + \omega_\chi)}{(1 + \omega_\phi)(1 + \omega_\chi)-\omega_{\phi\chi}^2} 
+ \frac{m \left(s_B(\tau) - \frac{\eta_\chi}{2 + d}\hat{s}_B(\tau)\right)(1 + \omega_\phi)}{(1 + \omega_\phi)(1 + \omega_\chi)-\omega_{\phi\chi}^2} 
+ \frac{n\left(s_F(\tau) - \frac{\eta_\psi}{1 + d}\hat{s}_F(\tau)\right)}{1 + \omega_\psi}
\Bigg],
\end{multline}
\begin{multline} \label{eq: lphichi}
l^{\phi\chi} = \frac{2}{d}
\frac{h^2 (1 + \omega_\phi)\omega_{\phi\chi} }{\left((1 + \omega_\phi)(1 + \omega_\chi )-\omega_{\phi\chi}^2\right)^2 (1 + \omega_\psi)} \times \\
\times \Bigg[\frac{\left(s_B(\tau) - \frac{\eta_\phi}{2 + d} \hat{s}_B(\tau)\right)\left( \omega_{\phi\chi}^2 + (1 + \omega_\phi) (1 + \omega_\chi)\right)}{(1 + \omega_\phi)\left((1 + \omega_\phi)(1 + \omega_\chi)-\omega_{\phi\chi}^2\right) } + 
\frac{2 \left(s_B(\tau) - \frac{\eta_\chi}{2 + d}\hat{s}_B(\tau)\right)(1 + \omega_\phi)}{(1 + \omega_\phi)(1 + \omega_\chi) -\omega_{\phi\chi}^2 } + \frac{\left(s_F(\tau) - \frac{\eta_\psi}{1 + d}\hat{s}_F(\tau)\right)}{1 + \omega_\psi}
\Bigg],
\end{multline}
where $v_d^{-1} = 2^{d+1} \pi^{\frac{d}{2}}\Gamma(\frac{d}{2})$. Define $S^{\phi/\chi}(\tau) = s_B(\tau) - \frac{\eta_{\phi/\chi}}{d+2}\hat{s}_B(\tau)$, $S^\psi(\tau) = s_F(\tau) - \frac{\eta_\psi}{d+1}\hat{s}_F(\tau)$. The thermal factors read
\begin{equation}
    s_B(\tau) = \frac{v_{d-1}}{v_d} \frac{d}{d-1} \frac{\tau}{2\pi} \sum_{n\in \mathbb{Z}}^{|\tau n|\leq 1} (1-(\tau n)^2)^{\frac{d-1}{2}}, \quad \hat{s}_B(\tau) = \frac{v_{d-1}}{v_d} \frac{d}{d-1} \frac{d+2}{d+1} \frac{\tau}{2\pi} \sum_{n\in \mathbb{Z}}^{|\tau n|\leq 1} (1-(\tau n)^2)^{\frac{d+1}{2}},
\end{equation}
\begin{equation}
s_F(\tau) = \frac{v_{d-1}}{v_d} \frac{d}{d-1}\frac{\tau}{2\pi} \sum_{n_F \in \mathbb{Z} + \frac{1}{2}}^{|\tau n_F|\leq 1} (1-\tau^2n_F^2)^{\frac{d-1}{2}},
\end{equation}
\begin{equation}
    \hat{s}_F(\tau) = \frac{v_{d-1}}{v_d} \frac{d (d+1)}{d-1}\frac{\tau}{2\pi} \sum_{n_F \in \mathbb{Z} + \frac{1}{2}}^{|\tau n_F|\leq 1} (1-\tau^2 n_F^2)^{\frac{d-1}{2}} \left(1-|\tau n_F|  \,
   _2F_1\left(-\frac{1}{2},\frac{d-1}{2};\frac{d+1}{2};1-\frac{1}{\tau^2 n_F^2}\right)\right), 
\end{equation}
\begin{equation}
    t_F(\tau) = \frac{v_{d-1}}{v_d}\frac{d}{2} \frac{d-2}{d-1}\frac{\tau }{2 \pi} \sum_{n_F \in \mathbb{Z} + \frac{1}{2}}^{|\tau n_F|\leq 1} \frac{\left(1-\tau ^2 n_F^2\right)^{\frac{d+1}{2}}}{\tau ^4 n_F^4} \left(\tau ^2 n_F^2-\frac{d-1}{d+1} \,
   _2F_1\left(1,\frac{d+1}{2};\frac{d+3}{2};1-\frac{1}{\tau ^2 n_F^2}\right)\right),
\end{equation}
\begin{equation}
    s_{BF} (\tau) = s_{F} (\tau), \quad \hat{t}_{BF}(\tau) = \frac{v_{d-1}}{v_d} \frac{d}{d-1} \frac{\tau}{2\pi} \sum_{n \in \mathbb{Z}}^{|\tau n_F|\leq 1}\frac{1}{|\tau n_F|} (1- \tau^2 n_F^2)^{\frac{d+1}{2}}  \,
   _2F_1\left(\frac{1}{2}, \frac{d+1}{2}, \frac{d+3}{2}, 1- \frac{1}{\tau^2 n_F^2}\right). 
\end{equation}
where $\tau = 2\pi T/k$, $n_F \equiv n + \tfrac{1}{2}$, and all the thermal factors are equal to one for $\tau \rightarrow 0$. For $\tau > 1$ in the bosonic thermal factors, only the zeroth Matsubara mode survives, while for $\tau > 2$ in the fermionic thermal factors, no mode survives and they evaluate to zero. 

To calculate the thermal effects for the fermionic anomalous dimension and the flow of the Yukawa coupling, where bosonic and fermionic Matsubara modes are mixed, we resort to the approximation introduced in \cite{Braun:2009ewx} and later employed in many works, e.g., \cite{Scherer:2012fjq, Tolosa-Simeon:2025fot}. In this approach, we set $\frac{1}{2}\tau \equiv \frac{\pi T}{k} \ll 1$, so that the bosonic Matsubara mode can be identified with the fermionic one, $n = n_F$. The replacement is applied to the $\hat{t}_{BF}$, $s_{BF}$ thermal factors, and to the $l$ threshold functions. To see why this approximation works well, consider RG scales $k < \pi T$. At such scales, the fermionic degrees of freedom do not affect the flow of the bosonic couplings, as due to the absence of zero fermionic Matsubara modes $S_\psi = 0$ in the flow of the effective potential (\ref{eq: uflow}). 

In the current work, numerical studies are performed at $T = 0$, and all the thermal factors are set to unity without approximations.

\section{Flow of the bosonic couplings}
The flow of the effective potential (\ref{eq: uflow}) is obtained by projecting the Wetterich equation (\ref{eq: Wetterich}) onto the constant $\phi$, $\chi$ field configurations with $\psi=\bar{\psi} = 0$

\begin{equation}
    \partial_t U_k[\rho_\phi, \rho_\chi] = \partial_t \Gamma_k [\Phi] \biggr|_{
    \begin{smallmatrix}
        \text{const } \phi, \chi\\
        \psi = \bar{\psi} = 0        
    \end{smallmatrix}
} = \frac{1}{2} \text{STr}\left(\left[\Gamma^{(2)}_k[\Phi]+R_k\right]^{-1} \partial_t R_k\right) \Biggr|_{
    \begin{smallmatrix}
        \text{const } \phi, \chi\\
        \psi = \bar{\psi} = 0        
    \end{smallmatrix}
}.
\end{equation}
The $\beta^{\text{FRG}}$-functions of the dimensionless couplings of the effective potential are obtained depending on the regime
\begin{enumerate}
    \item SYM-SYM: 
    \begin{equation}
        \partial_t \bar{\lambda}_{nm} = \left(\partial_t u\right)^{(n,m)}\Big|_{\begin{smallmatrix}
        \bar{\rho}_\phi = 0\\
        \bar{\rho}_\chi = 0
    \end{smallmatrix}},
    \end{equation}
    \item SYM-SSB:
    \begin{equation}
        \partial_t \bar{\lambda}_{nm} = (\partial_t u)^{(n,m)} + \bar{\lambda}_{n,m+1} \partial_t \bar{\kappa}_\chi\Big|_{\begin{smallmatrix}
        \bar{\rho}_\phi = 0\\
        \bar{\rho}_\chi = \bar{\kappa}_\chi
    \end{smallmatrix}}, \quad \partial_t \bar{\kappa}_\chi = - \frac{(\partial_t u)^{(0,1)}}{\bar{\lambda}_{0,2}}\Big|_{\begin{smallmatrix}
        \bar{\rho}_\phi = 0\\
        \bar{\rho}_\chi = \bar{\kappa}_\chi
    \end{smallmatrix}},
    \end{equation}
\item SSB-SYM:
    \begin{equation}
        \partial_t \bar{\lambda}_{nm} = (\partial_t u)^{(n,m)} + \bar{\lambda}_{n+1,m} \partial_t \bar{\kappa}_\phi\Big|_{\begin{smallmatrix}
        \bar{\rho}_\phi =\bar{\kappa}_\phi\\
        \bar{\rho}_\chi = 0
    \end{smallmatrix}}, \quad \partial_t \bar{\kappa}_\phi = - \frac{(\partial_t u)^{(1,0)}}{\bar{\lambda}_{2,0}}\Big|_{\begin{smallmatrix}
    \bar{\rho}_\phi = \bar{\kappa}_\phi\\
        \bar{\rho}_\chi = 0
    \end{smallmatrix}},
    \end{equation}
    \item SSB-SSB: 
    \begin{equation}
        \partial_t \bar{\lambda}_{nm} = (\partial_t u)^{(n,m)} + \bar{\lambda}_{n+1, m} \partial_t \bar{\kappa}_\phi + \bar{\lambda}_{n, m+1} \partial_t \bar{\kappa}_\chi \Big|_{\begin{smallmatrix}
    \bar{\rho}_\phi = \bar{\kappa}_\phi\\
        \bar{\rho}_\chi = \bar{\kappa}_\chi
    \end{smallmatrix}},
    \end{equation}
    \begin{equation}
        \partial_t \bar{\kappa}_\phi = \frac{\bar{\lambda}_{0,2} \left(\partial_t u\right)^{(1,0)} - \bar{\lambda}_{1,1} (\partial_t u)^{(0,1)}}{\bar{\lambda}_{1,1}^2 - \bar{\lambda}_{2,0} \bar{\lambda}_{0,2}}\Bigg|_{\begin{smallmatrix}
            \bar{\rho}_\phi = \bar{\kappa}_\phi\\
            \bar{\rho}_\chi = \bar{\kappa}_\chi
        \end{smallmatrix}}, \quad
        \partial_t \bar{\kappa}_\chi = \frac{\bar{\lambda}_{2,0} \left(\partial_t u\right)^{(0,1)} - \bar{\lambda}_{1,1} (\partial_t u)^{(1,0)}}{\bar{\lambda}_{1,1}^2 - \bar{\lambda}_{2,0} \bar{\lambda}_{0,2}}\Bigg|_{ \begin{smallmatrix}
            \bar{\rho}_\phi = \bar{\kappa}_\phi\\
            \bar{\rho}_\chi = \bar{\kappa}_\chi
        \end{smallmatrix}}.
    \end{equation}
\end{enumerate}

\section{Anomalous dimensions and the Yukawa coupling flow} \label{app: eta}

The anomalous dimension $\eta_\Phi = - \partial_t \ln Z_{\Phi, k}$ can be extracted from the flow of the exact inverse propagator $\Gamma^{(2)}_k$. We use the background field expansion $\phi^a = \delta^{a,1} \phi + \xi^a_\phi$, $\chi^b = \delta^{b, 1} \chi + \xi^b_\chi$, $\psi \equiv \xi_\psi$, $\bar{\psi} \equiv \xi_{\bar{\psi}}$ keeping $\phi$ and $\chi$ constant and varying one of the $\xi$ modes in the expression for the flow of the average action $\partial_t\Gamma_k[\Phi]$ (\ref{eq: Wetterich}) to obtain $\partial_t \Gamma_k^{(2)}$. From the momentum space form of the kinetic term  $\Gamma_k[\Phi] = \int \frac{d^d p}{(2\pi)^d} \frac{Z\phi}{2} p^2 \phi(p) \phi(-p) 
 + (\phi \leftrightarrow \chi) + Z_{\psi,k} \bar{\psi}(p) i \slashed{p} \psi(p) + \dots$ it follows that 
\begin{equation} \label{eq: Zflow}
    \partial_t Z_{\phi/\chi} = \left.\frac{1}{2d} \partial_{p_\mu} \partial_{p_\mu} \frac{\delta^2}{\delta \xi^a_{\phi/\chi}(-p) \delta \xi^a_{\phi/\chi}(p)} \partial_t \Gamma[\Phi]\right|_{\begin{smallmatrix}
         \xi_{\Phi} = 0\\
         p_\mu =0 
    \end{smallmatrix}}, \quad \partial_t Z_{\psi} = \left.\frac{1}{i d_\gamma d} \left(\gamma^\mu\right)_{\beta \alpha} \partial_{p_\mu} \frac{\vec{\delta}}{\delta \bar{\psi}_{\alpha}(p)} \partial_t \Gamma[\Phi] \frac{\cev{\delta}}{\delta \psi_{\beta}(p)}\right|_{\begin{smallmatrix}
         \xi_{\Phi} = 0\\
         p_\mu =0 
    \end{smallmatrix}},
\end{equation}
where summation over $\mu \in \{1,\dots,d\}$ and spinor indices $\alpha, \beta \in \{1,\dots,d_\gamma\}$ is performed, but $a$ is kept fixed. 

Using the Wetterich equation (\ref{eq: Wetterich}) and the $\tilde{\partial}_t$ derivative (\ref{eq: dttilde}), one obtains the flow of the second variation of the average action
\begin{multline} \label{eq: inverse_propagator_flow}
    \partial_t \Gamma^{(2)}_k = \frac{\vec{\delta}}{\delta \Phi(-p)} \partial_t\Gamma_k[\Phi] \frac{\cev{\delta}}{\delta \Phi(p)} = \frac{1}{2} \frac{\vec{\delta}}{\delta \Phi(-p)} \tilde{\partial}_t \text{STr} \ln \left(\Gamma^{(2)}_k+R_k\right) \frac{\cev{\delta}}{\delta \Phi(p)} =\\= -\frac{1}{2} \tilde{\partial}_t \text{STr} \left(\Gamma^{(2)}_k+R_k\right)^{-1} \vec{\Gamma}^{(3)}_{-p; \,k} \left(\Gamma^{(2)}_k+R_k\right)^{-1} \cev{\Gamma}^{(3)}_{p;\,k} +\frac{1}{2} \tilde{\partial}_t \text{STr} \left(\Gamma^{(2)}_k+R_k\right)^{-1} \Gamma^{(4)}_{-p,p; \,k},
\end{multline}
where 
\begin{equation}
(\Gamma^{(2)}_{k})_{q_1, q_2} = \frac{\vec{\delta}}{\delta \Phi(-q_1)}\Gamma_k[\Phi]\frac{\cev{\delta}}{\delta \Phi(q_2)}, \quad (\vec{\Gamma}^{(3)}_{p;\,k})_{q_1, q_2} = \frac{\vec{\delta}}{\delta \Phi(p)} (\Gamma^{(2)}_{k})_{q_1, q_2}, \quad (\Gamma^{(4)}_{-p,p;\,k})_{q_1, q_2} =  (\vec{\Gamma}^{(3)}_{-p;\,k})_{q_1, q_2}\frac{\cev{\delta}}{\delta \Phi(p)}    
\end{equation}
are the second, third, and fourth variations of the average action, respectively, with $\Phi(p) = (\phi(p), \chi(p), \psi(p), \bar{\psi}(-p))$; index structure of $\Phi$ is omitted.  Since for our purposes (\ref{eq: Zflow}) we set the fields to constant values after taking the variations, all the variations under the supertrace should satisfy momentum conservation separately, i.e. $(\Gamma^{(2)}_k)_{q_1,q_2} \sim \delta(-q_1+q_2)$, $(\Gamma^{(3)}_{p;k})_{q_1,q_2} \sim \delta(p-q_1+q_2)$, $(\Gamma^{(4)}_{-p,p;k})_{q_1,q_2} \sim \delta(-q_1+q_2)$. Thus, the term with $\Gamma^{(4)}$ in (\ref{eq: inverse_propagator_flow}) does not depend on $p_\mu$ and consequently does not contribute to the anomalous dimension.

We are using the prescription of projecting onto the transverse modes. According to it, for the case of scalar fields, the variation in (\ref{eq: Zflow}) is taken with respect to a transverse mode $\xi^{a\neq 1}$. For the $\phi$ field, it is straightforward to evaluate the prescription; however, the $\chi$ field is a single scalar and does not have transverse modes. To resolve this, the field $\chi$ is promoted to an $O(M)$ scalar, $M > 1$, with the Yukawa interaction $\sum_{j=1}^M h \chi_j \bar{\psi} \psi$. The expression of the anomalous dimension $\eta_\chi$ obtained this way does not depend on $M$, so at the end of the calculation, one sets back $M=1$. It was argued in \cite{Hawashin:2024dpp, Classen:2015mar} that for the calculation of the anomalous dimension, projecting onto a transverse mode, rather than the radial mode, yields better numerical results. See Appendix C of \cite{Classen:2015mar} for details. 

We list the expressions for the anomalous dimensions
\begin{equation} \label{eq: etaphi}
    \eta_\phi = \frac{16 v_d}{d} \frac{\bar{\rho}_\chi  \left(u^{(1,1)}\right)^2 \left((1+\omega_\phi)^2+\omega_{\phi \chi} ^2\right)-\omega_{\phi \chi}^2  u^{(2,0)} (2+\omega_\chi +\omega_\phi)+\bar{\rho}_\phi  \left(u^{(2,0)}\right)^2 \left((1+\omega_\chi)^2+\omega_{\phi \chi}
   ^2\right)}{\left(1+u^{(1,0)}\right)^2 \left((1+\omega_\chi)(1+\omega_\phi) -\omega_{\phi \chi}^2\right)^2} s_B(\tau),
\end{equation}
\begin{multline} \label{eq: etachi}
    \eta_\chi = \frac{16 v_d}{d} \frac{\bar{\rho}_\phi  \left(u^{(1,1)}\right)^2
   \left((1+\omega_\chi)^2+\omega_{\phi \chi} ^2\right)-\omega_{\phi \chi} ^2 u^{(0,2)} (2+\omega_\chi +\omega_\phi) + \bar{\rho}_\chi  \left(u^{(0,2)}\right)^2 \left((1+\omega_\phi )^2+\omega_{\phi \chi} ^2\right)}{
   \left(1+u^{(0,1)}\right)^2 \left((1+\omega_\chi) (1+\omega_\phi)-\omega_{\phi \chi} ^2\right)^2} s_B(\tau) + \\ +\frac{8 v_d}{d} N_2 \bar{h}^2 m_{F} (\omega_\psi,\eta_\psi, \tau) ,
\end{multline}
where $\eta_\chi$ is symmetric to $\eta_\phi$ expression with the addition from the Yukawa-interaction. For the fermion anomalous dimension, we have 
\begin{equation} \label{eq: etapsi}
    \eta_\psi = \frac{8 v_d h^2}{d} \frac{\left(s_{BF}(\tau)-\frac{\eta_\phi }{d+1} \hat{t}_{BF}(\tau)\right) \omega_{\phi \chi} ^2+\left(s_{BF}(\tau)-\frac{\eta_\chi }{d+1} \hat{t}_{BF}(\tau)\right) (1+\omega_\phi )^2}{(1+\omega_\psi )
   \left((1+\omega_\chi ) (1+\omega_\phi )-\omega_{\phi \chi} ^2\right)^2}. 
\end{equation}
From the form of the Yukawa coupling in the average effective action $\Gamma_k[\Phi] = \dots +h\chi\bar{\psi} \psi +\dots$ it follows that 
\begin{equation}
    \partial_t h = \frac{1}{d_\gamma}\frac{\delta}{\delta \chi} \left.\frac{\vec{\delta}}{\delta \bar{\psi}_{\alpha}(p)} \partial_t \Gamma[\Phi] \frac{\cev{\delta}}{\delta \psi_{\alpha} (p)}\right|_{\tiny
         \xi_{\Phi} = 0,\,\, p = 0}.
\end{equation}
Using the Wetterich equation (\ref{eq: Wetterich}) and similar to (\ref{eq: inverse_propagator_flow}) derivation, one obtains (\ref{eq: hflow}).

\section{Comparison of fixed point data with recent studies}

In this work, we construct an RG trajectory connecting two CFTs to demonstrate persistent spontaneous parity breaking. Obtaining high-precision critical exponents and anomalous dimensions of these CFTs is beyond our scope. However, to support the choice of our truncation scheme for establishing the main results, we include Table~\ref{tab: CFTs_comparison}, which compares our fixed-point data with recent developments in the field.
\begin{table}[h]
  \centering
  \setlength{\arrayrulewidth}{0.4pt}
  \renewcommand{\arraystretch}{1.2}
  \begin{tabular}{|l c c|l c c c|}
    \hline
    \multicolumn{3}{|c|}{$O(N)$ critical model ($N=100$)} 
      & \multicolumn{4}{c|}{Gross–Neveu–Yukawa critical model ($N_f d_\gamma=8$)} \\
    \hline
    Study & $1/\nu$ & $\eta_\phi$ 
      & Study & $1/\nu$ & $\eta_{\chi}$ & $\eta_{\psi}$ \\
    \hline
    This work, FRG [LPA$'$12, $R_k^{\text{lin}}$]
      & 1.0102   & 0.00247    
      & This work, FRG [LPA$'$12, $R_k^{\text{lin}}$] 
      & 0.982    & 0.760     & 0.032    \\
    $1/N$-expansion ($1^\text{st}$/$2^\text{nd}$ order) \cite{Moshe:2003xn}  & 1.0108  & 0.00268    
      & QMC~\cite{PhysRevB.108.L121112} 
      & 1.07(12) & 0.72(6)   & 0.04(2)  \\
       FRG $O(\partial^4)$~\cite{DePolsi:2020pjk} 
      & 1.0113(2)   & 0.00268(4)
      & $\epsilon$‑expansion (4 loops)~\cite{PhysRevB.98.125109}
      & 0.993(27)& 0.704(15)& 0.043(12)\\
    \multicolumn{3}{|c|}{} 
      & Conformal bootstrap~\cite{Erramilli:2022kgp}
      & 0.998(12)& 0.7329(27)&0.04238(11)\\
    \hline
  \end{tabular}
  \caption{Critical exponents ($1/\nu = \theta$) and anomalous dimensions of $O(N)$ and Gross-Neveu-Yukawa critical models in $d=3$. }
  \label{tab: CFTs_comparison}
\end{table}

\noindent
For the critical biconical model, we adopt the truncation scheme as in recent work~\cite{Hawashin:2024dpp}, where fixed-point data was compared to the 5-loop $\epsilon$-expansion for $N_1=2$. In the limit $N_1 \rightarrow \infty$, the LPA becomes exact, and for large finite values of $N_1$, the study demonstrates convergence of the model within LPA$'$ and LPA.

\twocolumngrid  

\end{document}